\documentclass[times,twocolumn,review]{elsarticle}

\usepackage{medima}
\usepackage{framed,multirow}
\usepackage{graphicx}
\usepackage{subcaption} 
\usepackage{textcomp}
\usepackage{blindtext}

\usepackage[namelimits]{amsmath} 
\usepackage{amssymb}             
\usepackage{amsfonts}            
\usepackage{mathrsfs}            
\usepackage{latexsym}
\usepackage{booktabs}
\usepackage{lineno}
\modulolinenumbers[5]
\usepackage{lineno,hyperref}

\usepackage{url}
\usepackage[table,xcdraw]{xcolor}
\usepackage{hyperref}
\usepackage{makecell}

\begin{document}

\verso{Gongning Luo \textit{et~al.}}

\begin{frontmatter}

\title{Tumor Detection, Segmentation and Classification Challenge on Automated 3D Breast Ultrasound: The TDSC-ABUS Challenge
}%




\author[1]{Gongning {Luo}}
\author[1]{Mingwang Xu}
\author[2]{Hongyu Chen}
\author[1]{Xinjie Liang}
\author[3,4]{Xing Tao}
\author[3,4,5]{Dong Ni}
\author[6]{Hyunsu Jeong}
\author[6]{Chulhong Kim}
\author[7,8]{Raphael Stock}
\author[7,8,9]{Michael Baumgartner}
\author[7,8,10]{Yannick Kirchhoff}
\author[7,8]{Maximilian Rokuss}
\author[7,11]{Klaus Maier-Hein}
\author[12]{Zhikai Yang}
\author[12]{Tianyu Fan}
\author[13]{Nicolas Boutry}
\author[13]{Dmitry Tereshchenko}
\author[13]{Arthur Moine}
\author[13]{Maximilien Charmetant}
\author[14]{Jan Sauer}
\author[14,15]{Hao Du}
\author[16]{Xiang-Hui Bai}
\author[16]{Vipul Pai Raikar}
\author[17]{Ricardo Montoya-del-Angel}
\author[17]{Robert Martí}
\author[18]{Miguel Luna}
\author[19]{Dongmin Lee}
\author[20]{Abdul Qayyum}
\author[21]{Moona Mazher}
\author[22]{Qihui Guo}
\author[22]{Changyan Wang}
\author[23,24]{Navchetan Awasthi}
\author[25]{Qiaochu Zhao}
\cortext[cor1]{Corresponding author:}
\author[1]{Wei {Wang}\corref{cor1}}
\ead{wangwei2019@hit.edu.cn}
\author[1]{Kuanquan {Wang}\corref{cor1}}
\ead{wangkq@hit.edu.cn}
\author[26]{Qiucheng {Wang}\corref{cor1}}
\ead{wangkq@hit.edu.cn}
\author[27]{Suyu {Dong}\corref{cor1}}
\ead{dongsuyu@nefu.edu.cn}

\address[1]{School of Computer Science and Technology, Harbin Institute of Technology, Harbin 150001, China.}
\address[2]{Department of Mathematics, Faculty of Science, National University of Singapore, Singapore.}
\address[3]{National-Regional Key Technology Engineering Laboratory for Medical Ultrasound, School of Biomedical Engineering, Shenzhen University Medical School, Shenzhen University, Shenzhen, China.}
\address[4]{Medical Ultrasound Image Computing (MUSIC) Laboratory, Shenzhen University, Shenzhen, China.}
\address[5]{School of Biomedical Engineering and Informatics, Nanjing Medical University, Nanjing, China.}
\address[6]{Departments of Electrical Engineering, Convergence IT Engineering, Mechanical Engineering, Medical Science and Engineering, Graduate School of Artificial Intelligence, and Medical Device Innovation Center, Pohang University of Science and Technology (POSTECH), Pohang, Republic of Korea.}
\address[7]{German Cancer Research Center (DKFZ) Heidelberg, Division of Medical Image Computing, Heidelberg, Germany.}
\address[8]{Faculty of Mathematics and Computer Science, Heidelberg University, Germany.}
\address[9]{Helmholtz Imaging, German Cancer Research Center (DKFZ), Heidelberg, Germany.}
\address[10]{HIDSS4Health - Helmholtz Information and Data Science School for Health, Karlsruhe/Heidelberg, Germany.}
\address[11]{Pattern Analysis and Learning Group, Department of Radiation Oncology, Heidelberg University Hospital.}
\address[12]{Department of Biomedical Engineering and Health, KTH Royal Institute of Technology, Stockholm, Sweden.}
\address[13]{EPITA Research Laboratory (LRE).}
\address[14]{FathomX.}
\address[15]{Saw Swee Hock School of Public Health, National University of Singapore.}
\address[16]{Philips Research.}
\address[17]{Computer Vision and Robotics Institute (ViCOROB), University of Girona.}
\address[18]{Department of Robotics and Mechatronics Engineering, DGIST, Korea.}
\address[19]{Department of Interdisciplinary Studies of Artificial Intelligence, DGIST, Korea.}
\address[20]{National Heart and Lung Institute, Faculty of Medicine, Imperial College London, London, United Kingdom.}
\address[21]{Centre for Medical Image Computing, Department of Computer Science, University College London, London, United Kingdom.}
\address[22]{The SMART (Smart Medicine and AI-based Radiology Technology) Lab, School of Communication and Information Engineering, Shanghai University, Shanghai, China.}
\address[23]{Faculty of Science, Mathematics and Computer Science, Informatics Institute, University of Amsterdam, Amsterdam, 1090 GH, The Netherlands.}
\address[24]{Department of Biomedical Engineering and Physics, Amsterdam UMC, Amsterdam, 1081 HV, The Netherlands.}
\address[25]{Xi’an Jiaotong-Liverpool University.}
\address[26]{Department of Ultrasound, Harbin Medical University Cancer Hospital, No. 150, Haping Road, Nangang District, Harbin, Heilongjiang Province, China.}
\address[27]{College of computer and control engineering, Northeast Forestry University, Harbin, China.}

\received{xx xx 202x}
\finalform{xx xx 202x}
\accepted{xx xx 202x}
\availableonline{xx xx 202x}

\begin{abstract}

Breast cancer is one of the most common causes of death among women worldwide. Early detection helps in reducing the number of deaths. Automated 3D Breast Ultrasound (ABUS) is a newer approach for breast screening, which has many advantages over handheld mammography such as safety, speed, and higher detection rate of breast cancer.
Tumor detection, segmentation, and classification are key components in the analysis of medical images, especially challenging in the context of 3D ABUS due to the significant variability in tumor size and shape, unclear tumor boundaries, and a low signal-to-noise ratio. The lack of publicly accessible, well-labeled ABUS datasets further hinders the advancement of systems for breast tumor analysis.
Addressing this gap, we have organized the inaugural \textbf{T}umor \textbf{D}etection, \textbf{S}egmentation, and \textbf{C}lassification Challenge on \textbf{A}utomated 3D \textbf{B}reast \textbf{U}ltra\textbf{s}ound 2023 (TDSC-ABUS2023). This initiative aims to spearhead research in this field and create a definitive benchmark for tasks associated with 3D ABUS image analysis. 
In this paper, we summarize the top-performing algorithms from the challenge and provide critical analysis for ABUS image examination. We offer the TDSC-ABUS challenge as an open-access platform at \url{https://tdsc-abus2023.grand-challenge.org/} to benchmark and inspire future developments in algorithmic research.

\end{abstract}

\begin{keyword}
\KWD Segmentation\sep Pulmonary artery\sep Multi-level\sep Efficiency
\end{keyword}

\end{frontmatter}

 
\section{Introduction}
\label{introduction}
Breast cancer, the most commonly diagnosed cancer worldwide, is also the fifth leading cause of death globally \citep{sung2021global}. The past decade has seen significant strides in reducing mortality and improving the 5-year survival rate, largely thanks to advancements in early detection and diagnostics \citep{kaye2002new}. However, accurately diagnosing malignant breast tumors at an early stage remains a paramount yet challenging goal.-

Central to breast cancer screening and diagnosis is the use of ultrasound technology. Traditional 2D hand-held ultrasound systems (HHUS), despite their widespread use, are hampered by operator dependency, time-consuming processes, and their limitation to two-dimensional imaging \citep{boca2021pros}. The Automated Breast Ultrasound System (ABUS) represents a groundbreaking advancement in this field. It not only standardizes the scanning process but also minimizes the influence of the operator on the quality of the images. By providing a detailed 3D representation of breast tissue, ABUS enables multi-angle evaluations and enriches the potential for retrospective analyses, thereby outperforming traditional ultrasound in tumor detection \citep{kaplan2014automated, zhang2012detection, xiao2015efficacy}. Nonetheless, the intricate interpretation of ABUS images requires extensive clinical experience, and current techniques have yet to fully harness the wealth of data offered by 3D imaging. Thus, there is a pressing need for the development of more advanced and efficient computer-aided diagnosis (CAD) algorithms for use in clinical settings.

In the realm of CAD, tumor detection, segmentation, and classification are three fundamental tasks, each serving as a critical preliminary step for further analysis. Numerous studies have been conducted to address these challenges across various medical contexts \citep{hwang2016novel, jimeno2022artifactid, albawi2023skin, alshmrani2023deep, gupta2023deep, gauriau2023head, hasan2023fp, zhang2024deep}. Focusing on ABUS images, significant progress has been made. For instance, in tumor detection, \cite{muramatsu2018mass} proposed a novel approach using a GoogLeNet-based fully convolutional network, leveraging both axial and reconstructed sagittal slices \citep{szegedy2016rethinking}. This method, while effective, highlighted the need for a more comprehensive approach to spatial information integration. In response, \cite{wang2018densely} developed a 3D U-net-based network that directly processes 3D ABUS images, thus overcoming the limitations of previous methods and enhancing diagnostic accuracy.

The segmentation of tumors in ABUS images is particularly challenging due to indistinct lesion boundaries. Addressing this, \cite{fayyaz2021mass} innovated a dual-path U-net architecture, achieving promising results. Moreover, \cite{zhou2021cross} further advanced segmentation techniques by devising a cross-model attention-guided network, incorporating an improved 3D Mask R-CNN head into a V-Net framework. Their methodology showed marked improvements over existing segmentation approaches.

Finally, in the field of tumor classification, \cite{zhuang2021tumor} introduced a shallowly dilated convolutional branch network (SDCB-Net), specifically designed for classifying breast tumors in ABUS images. Additionally, \cite{kim2022mask} proposed an innovative branch network that integrates segmenting mask information into the training regime, noting that the performance varied depending on the positioning of the mask branch network in relation to mass and cancer classifications.
\begin{figure}[htbp]
	\centering
	\includegraphics[width=\columnwidth]{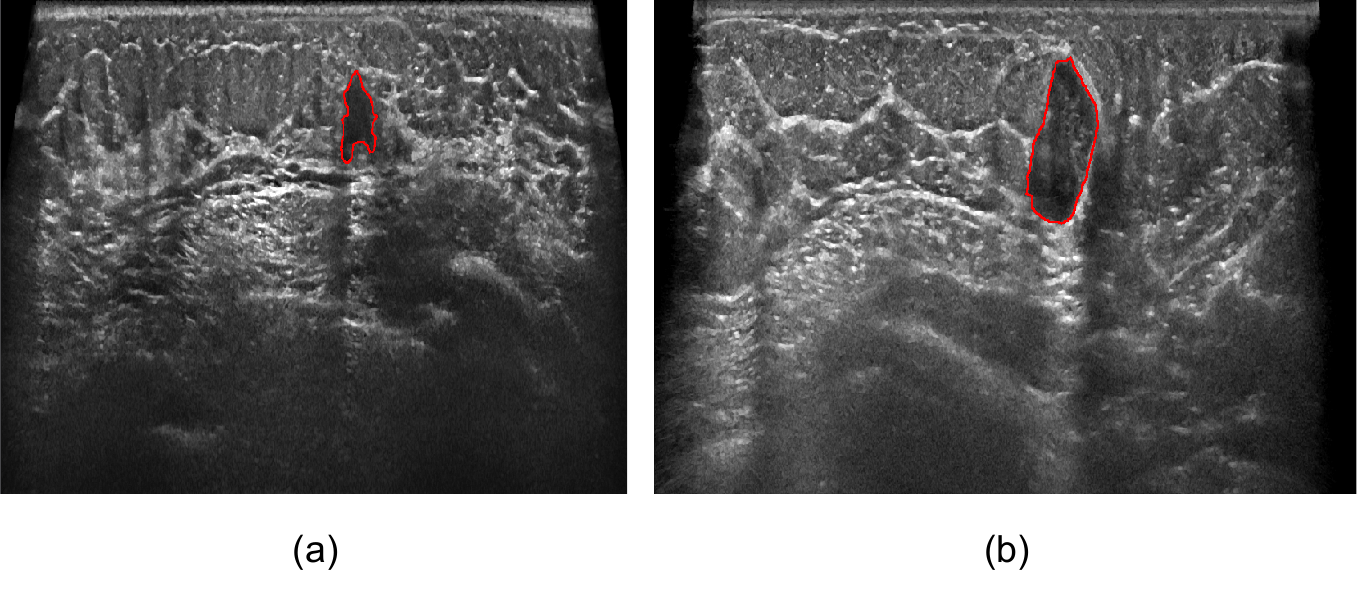}
	\caption{Representative ABUS Image Cases. (a) A small-sized tumor exhibiting a jagged boundary. (b) A comparatively larger tumor characterized by a smooth periphery.}
  \label{fig:example}
\end{figure}

Despite the advancements in 3D Automated Breast Ultrasound (ABUS) imaging, developing efficient and accurate CAD systems remains a significant challenge. As shown in Fig. \ref{fig:example}, although ABUS addresses some limitations of 2D ultrasound, it still presents distinct challenges for CAD systems:
(1) \textbf{Unclear Boundaries:} 
The inherent characteristics of ultrasound imaging often result in blurred lesion boundaries, complicating the tasks of annotation and segmentation.
(2) \textbf{High Similarity Between Lesion and Background:}
Lesions may exhibit features akin to the surrounding tissue, leading to potential misclassification as tumors or, conversely, missed detections.
(3) \textbf{Wide Variation in Lesion Size and Shape:}
The size of lesions can vary considerably, posing challenges for algorithms to effectively adapt to extreme cases. Additionally, the shape of tumors can range from smooth to rough textures, making it difficult to differentiate based on these characteristics alone.
(4) \textbf{Small Proportion of Lesion in the Total Image:}
Lesions often occupy a minor fraction of the overall image area, which can make detection challenging due to the predominance of non-lesion tissue.
(5) \textbf{Unpredictable Lesion Location:}
The location of lesions within the image can be highly variable, adding complexity to the detection process as there is no standard region to focus the analysis.

Over recent decades, despite extensive research into Automated Breast Ultrasound (ABUS) imaging, there remains a notable absence of a publicly available ABUS dataset that serves as a standard benchmark for the fair evaluation of algorithms.

To bridge this gap, we have initiated the \textbf{T}umor \textbf{D}etection, \textbf{S}egmentation, and \textbf{C}lassification Challenge on \textbf{A}utomated 3D \textbf{B}reast \textbf{U}ltra\textbf{s}ound 2023 (TDSC-ABUS2023). This challenge was held as part of the International Conference on Medical Image Computing and Computer Assisted Intervention (MICCAI) 2023, in Turkey. The competition invited participants to develop algorithms that address detection, segmentation, or classification of tumors in ABUS images. Entrants had the flexibility to tackle any combination of these tasks, with each task being independently evaluated using specific metrics. In order to promote a comprehensive approach, we also introduced an overall leaderboard encompassing all three tasks.

This paper presents a comprehensive overview of the TDSC-ABUS 2023 challenge, highlighting the most effective algorithms. Our primary contributions can be encapsulated as follows:
\begin{itemize}
  \item We have established the inaugural TDSC-ABUS challenge, the first of its kind to focus on the trifecta of detection, segmentation, and classification tasks in ABUS imaging, thereby offering a pioneering benchmark for ABUS CAD algorithm assessment.
  \item We provide a detailed analysis of the algorithms submitted, emphasizing the strategies and results of the leading teams.
  \item We delineate a range of algorithmic approaches tailored for the three fundamental tasks in ABUS imaging and offer insightful recommendations based on our findings.
\end{itemize}


\section{Methods}
\label{methods}
\subsection{Challenge dataset}

The TDSC-ABUS challenge provided a dataset comprising 200 Automated Breast Ultrasound (ABUS) images sourced from the Harbin Medical University Cancer Hospital. These images were precisely annotated by an experienced radiologist to delineate tumor boundaries and to categorize each case as malignant or benign. Furthermore, we derived tumor bounding boxes from these segmentation boundaries. The dimensions of the images vary, ranging from \(843 \times 546 \times 270\) to \(865 \times 682 \times 354\) pixels. The dataset features a pixel spacing of \(0.200\) mm and \(0.073\) mm, with the interslice spacing approximated at \(0.475674\) mm. Prior to their inclusion in the study, all data were anonymized to ensure privacy and compliance with ethical standards. The distribution of malignant to benign cases within the dataset is approximately \(0.58:0.42\), leading to a stratified sampling strategy for the division of training, validation, and testing subsets to mirror the real-world prevalence of these conditions. This methodology and the corresponding distribution details are depicted in Table \ref{tab:data}.

\begin{table}[h]
  \centering
  \caption{Dataset division details.}
  \label{tab:data}
  \setlength{\tabcolsep}{1.5mm}
  \begin{tabular}{cccc}
    \hline \toprule
    &Samples&Malignant&Benign\\
    \hline
    Training&100&58&42\\
    Validation&30&17&13\\
    Test&70&40&30\\
    \hline \toprule
  \end{tabular}
\end{table}


To ensure the accuracy and reliability of the annotations, ten experts, each boasting over five years of clinical experience, were enlisted for this task. They were organized into two groups of five, with each image receiving annotations from one group to foster a collaborative and comprehensive evaluation. The manual annotation process was conducted using the Seg3D software, followed by a thorough mutual review of the results to ensure consistency and accuracy.

The dataset was partitioned into three distinct subsets for the challenge: the training set, the validation set, and the testing set, containing 100, 30, and 70 cases, respectively. Challenge participants were granted access to the training images along with their annotations, and to the validation images without annotations, to simulate a realistic scenario for algorithm development and testing. The annotations for the validation set and the testing images, along with their annotations, were retained by the challenge organizers until the evaluation phase.

\subsection{Challenge organization}

In the organization of the TDSC-ABUS challenge and the drafting of this manuscript, we adhered to the guidelines as proposed by Maier et al. \cite{maier2020bias,maier2018rankings}. The structure of the TDSC-ABUS challenge was systematically divided into three phases: training, validation, and testing.

During the \textbf{training phase}, upon approval of their applications, participants were provided with 100 training images along with comprehensive annotations, which included segmentation boundaries, tumor categorizations, and detection boxes. This phase spanned approximately three months, offering participants ample time to develop and refine their algorithms for any of the challenge tasks. Recognizing the complexity of controlling the use of pre-trained models, participants were permitted to incorporate external datasets into their development process to enhance their algorithm's performance. Our overarching aim was to establish a current benchmark for the field.

The \textbf{validation phase}, lasting 46 days, presented participants with the validation dataset without labels, allowing them to submit their task results to the challenge's official website up to three times per day. The platform automatically displayed scores, enabling participants to gauge the efficacy of their algorithms in real-time.

In the \textbf{testing phase}, participants were required to encapsulate their algorithms within a Docker container, designed to function akin to a black box. These containers were then submitted to us for execution in a standardized environment. We provided feedback on the execution outcomes (successful, failed, or incorrect format) to the participants. A Docker container that successfully executed and yielded logical results was accepted as the team's final submission. The outcomes of these runs, encompassing scores for each specific case, were disclosed to the public. Teams with successful submissions were requested to submit a brief paper detailing their methodology. 

Recognition and awards were conferred upon all ranking teams, with the top team from each leaderboard receiving a cash prize. Teams that ranked successfully were also invited to present their innovative algorithms at the MICCAI challenges conference and to co-author the challenge review paper.

\subsection{Evaluation metrics and ranking scheme}
To ensure a thorough and precise evaluation of the algorithms, we employed a set of widely recognized metrics tailored to the specific demands of segmentation and classification tasks. For the classification task, we utilized Accuracy (ACC) and the Area Under the Receiver Operating Characteristic Curve (AUC) to measure the efficacy and reliability of the algorithms. In the task of segmentation, we applied the Dice Similarity Coefficient (Dice) and the Hausdorff Distance (HD) to assess the precision of the algorithm in identifying and delineating the target structures accurately.

Given the heightened challenge associated with the detection task—stemming from the fact that tumors often represent a minor fraction of the overall image area, we chose to implement the Free-Response Receiver Operating Characteristic (FROC) curve as the sole metric for this task. The FROC curve is particularly suited to this context as it offers a balanced evaluation of sensitivity against the rate of false positives, with the algorithm's performance delineated through the average sensitivity across a spectrum of false positive levels (FP = 0.125, 0.25, 0.5, 1, 2, 4, 8). Within this framework, a detection is classified as successful (a "hit") if the Intersection over Union (IoU) between the proposed bounding box and the ground-truth bounding boxes of the tumor exceeds a threshold of 0.3, ensuring that only precise detections are recognized.

After we got all the teams' final scores, we implemented the following ranking scheme:
\begin{itemize}
  \item Step 1. For each specific task, the initial step involved normalizing the scores for each team. This normalization was achieved through min-max scaling applied to the scores of teams with valid results, thereby standardizing the scores across a uniform scale.
  \item Step 2. Subsequently, we computed the average of these normalized scores for all teams within each task and arranged them in descending order to establish the leaderboard for each individual task. Notably, for the segmentation task, which utilizes the Hausdorff Distance (HD) metric where a lower score indicates better performance, we calculated the average score using the formula $(1 + Norm\_DICE - Norm\_HD) / 2$.
  \item Step 3. Our next step focused on identifying teams that participated across all three tasks. For these teams, we performed normalization again for each metric to ensure fairness across tasks.
  \item Step 4. Leveraging the newly normalized scores across all metrics, we calculated a comprehensive average score for each team using the formula $(1 + Norm\_DICE - Norm\_HD) / 2 + (Norm\_ACC + Norm\_AUC) / 2 + Norm\_FROC$. Teams were then ranked based on these average scores, from highest to lowest, to determine the overall leaderboard standings.

\end{itemize}

\section{Results}
\subsection{Challenge participants and submissions}
\begin{figure}
	\centering
	\includegraphics[width=\linewidth]{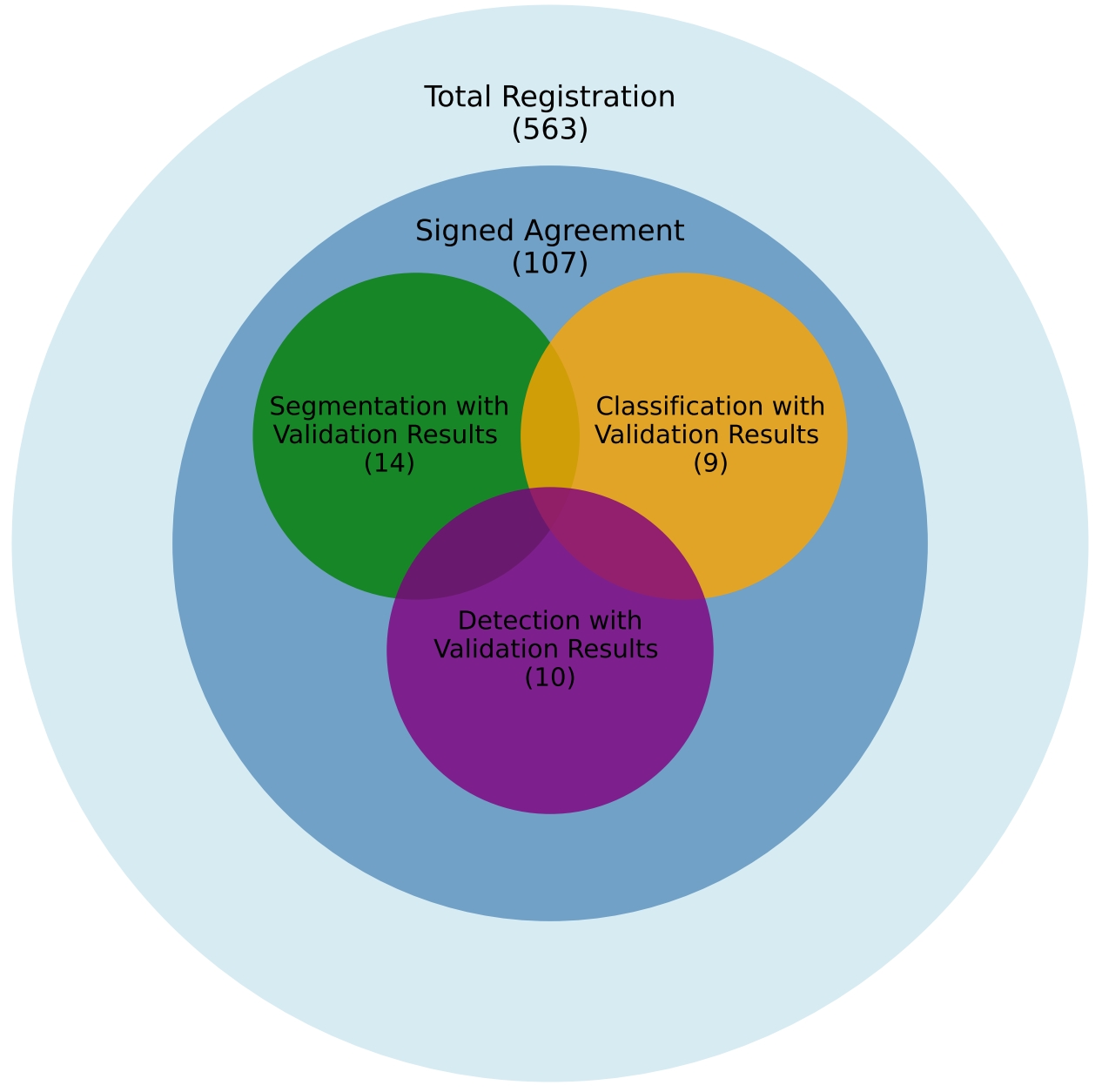}
	\caption[]{Summary of TDSC-ABUS 2023 Challenge participants and submissions. There were 563 teams registering on the official grand-challenge webpage and 107 of them signed the agreement. Finally, 14 teams submitted segmentation validation results, 9 teams submitted classification validation results, and 10 teams submitted detection validation results.}
	\label{Participants}
\end{figure}

The official TDSC-ABUS challenge was hosted on the grand-challenge platform\footnote{\url{https://tdsc-abus2023.grand-challenge.org/}}. Participants should sign the challenge rule agreement and send it to the official mailbox. Fig. \ref{Participants} shows information about participants and submissions. Specifically, we received more than 560 applications from over 49 countries and region on the grand-challenge webpage and 106 teams were approved with a complete signed application form. During the validation phase, we received 145 submissions for classification task, 398 submissions for segmentation task and 89 submissions for detection tasks. During the testing phase, 21 teams submitted Docker containers. But 3 Docker containers fail to run and 1 docker produceed no output. Finally, we got 17 qualified submissions.

\subsection{Algorithms summarization of teams}
We have 17 teams that produced valid results, but one of them did not submit short paper describe their method. Thus we summarize key point of other 16 teams in this section. Some teams solve all three tasks, while others not. Thus we highlight key points of each team algorithm of each task in Table \ref{tab:top10}. Their key strategies were demonstrated as follows.

\begin{table*}[ht]
  \centering
  \caption{Summary of the benchmark methods of top ten teams.}
  \label{tab:top10}
  \setlength{\tabcolsep}{3mm} 
  \renewcommand{\arraystretch}{1.2} 
  
  \begin{tabular}[]{p{1.5cm}p{5cm}p{4.5cm}p{4.5cm}}
    \hline
    \toprule
    Team & Detection & Classification & Segmentation \\ 
    \hline
    SZU & DETNet & CLSNet & SEGNet \\ \hline
    POSTECH & \makecell[l]{Segmentation-based detection \\ with Softmax} & \makecell[l]{nnU-Net with \\ Auxiliary classifier} & \makecell[l]{Multi-task 3D nnU-Net \\ with residual blocks} \\ \hline
    HU & nnDetection & ResNet-18 & nnU-Net \\ \hline
    KTH & Largest connected component & DenseNet201 & STU-NET \\ \hline
    EPITA & Bounding box from segmentation results & Size-based volume classification & 3D LowRes nnU-Net \\ \hline
    FX & \makecell[l]{Segmentation-based detection \\ with CRF and confidence filtering} & DenseNet121 & \makecell[l]{Swin-UNETR \\ and SegResNetVAE ensemble} \\ \hline
    PR & YOLOV8 & YOLOV8, MedSAM & MedSAM \\ \hline
    NV & - & - & SegResNet (Auto3DSeg framework) \\ \hline
    UDG & Bounding box from SAMed segmentation & - & SAMed (Segment Anything Model for medical images) \\ \hline
    DGIST & - & 2D U-Net classification branch & 2D U-Net \\ \hline
  \end{tabular}
\end{table*}

\subsubsection{T1: Shenzhen University (SZU)}

\begin{figure}[h]
    \centering
    \includegraphics[width=\columnwidth]{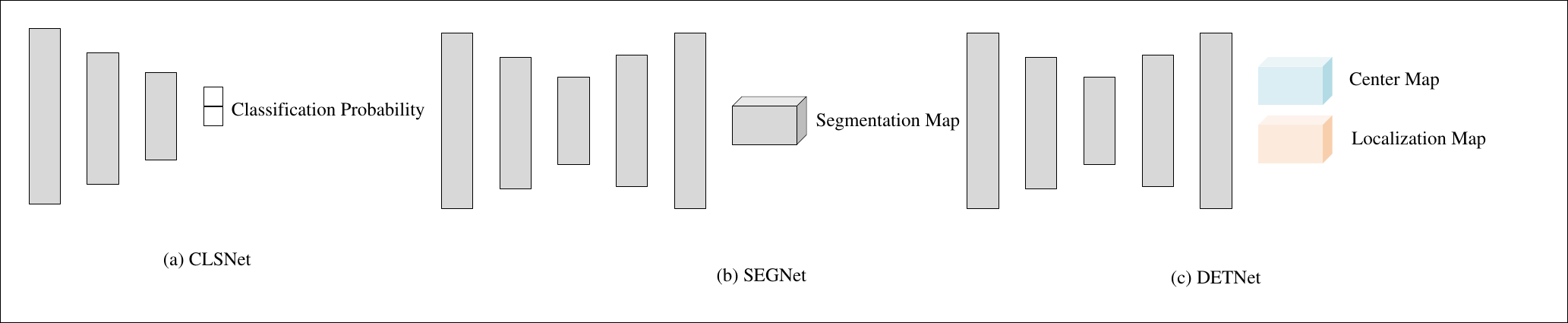} 
    \caption{The model structure of T1}
    \label{fig:model_structure_t12}
\end{figure}

SZU developed a unified framework for breast lesion classification, segmentation, and detection by designing three specialized models: CLSNet, SEGNet, and DETNet, each tailored for their respective tasks. An illustrative overview of the structures of these networks is depicted in Fig. \ref{fig:model_structure_t12}.

For classification, the team used a 3D ResNet architecture in CLSNet \citep{he_deep_2016}, which leverages softmax activation to differentiate between benign and malignant lesions. The 3D ResNet's ability to capture complex spatial patterns in breast ultrasound images enhances the model's precision in lesion characterization.

For segmentation, SZU built SEGNet on the traditional UNet architecture \citep{navab_u-net_2015}, which incorporates multi-scale features and skip connections between the encoder and decoder. This setup has been proven effective in medical image segmentation, enabling SEGNet to achieve accurate and contextually coherent delineation of breast lesions.

For detection, DETNet was inspired by the CPMNet approach \citep{martel_cpm-net_2020}, employing a single-stage anchor-free paradigm. Unlike multi-stage or anchor-based methods, DETNet offers improved inference efficiency and is better suited for handling a wide range of lesion sizes and characteristics, enhancing its ability to generalize across diverse cases.

SZU’s framework integrates these specialized models to address the distinct challenges of classification, segmentation, and detection, providing a comprehensive approach to 3D breast ultrasound analysis in breast cancer diagnosis.

\subsubsection{T2: Pohang University of Science and Technology (POSTECH)}
\begin{figure}[h]
    \centering
    \includegraphics[width=\columnwidth]{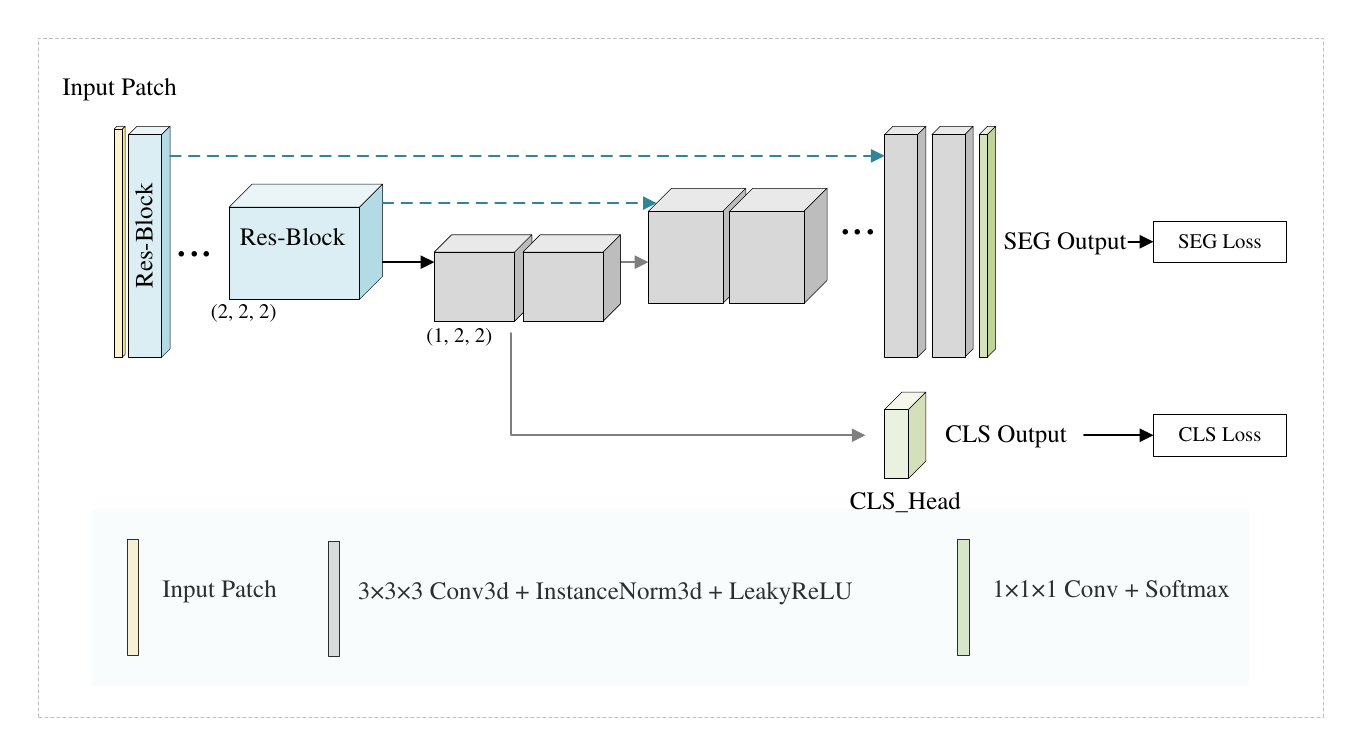} 
    \caption{The model structure of T2}
    \label{fig:model_structure_t10}
\end{figure}

POSTECH proposed a multi-task residual network for lesion detection in 3D breast ultrasound images. Based on the 3D nn-UNet architecture \citep{isensee_nnu-net_2021}, they enhanced the encoder with five residual blocks and adopted a multi-task learning strategy. This approach first performs segmentation, then uses the segmentation results to facilitate lesion detection. By incorporating segmentation and detection tasks, the network effectively identifies tumor regions with improved accuracy. The model structure of them is shown in Fig. \ref{fig:model_structure_t10}.

To address the class imbalance between lesion and non-lesion areas in ABUS data, POSTECH employed a patch-based training strategy. In this approach, positive patches containing lesions and random negative patches were selected as inputs to the network, allowing the model to focus more effectively on the small lesion areas.

The multi-task learning framework uses 3D nn-UNet as the backbone for both segmentation and detection. For segmentation, a binary network was trained with labels ‘1’ for lesions and ‘0’ for background, with skip connections added in the encoder to form residual blocks, enhancing feature transfer. After segmentation, Softmax was applied to calculate lesion probabilities and generate a 3D bounding box for detection. In the inference phase, sliding window inference and soft voting were employed to further refine classification outcomes, resulting in improved differentiation between benign and malignant nodules. This architecture, including residual blocks, effectively mitigates the vanishing gradient problem and improves generalization \citep{he_deep_2016}.

\subsubsection{ T3: Heidelberg University(HU)}
HU developed an advanced approach for 3D breast tumor detection, segmentation, and classification using state-of-the-art frameworks and careful methodical adjustments. For segmentation, the team utilized nnU-Net \citep{isensee_nnu-net_2021}, a well-established tool in medical imaging tasks, and adapted it to their dataset by processing full-resolution 3D images with a batch size of 2 and a patch size of [80, 160, 192], normalized using z-score. Training was carried out over 1000 epochs, and the model’s performance was evaluated with 5-fold cross-validation, using the Dice coefficient to measure accuracy. To enhance performance, HU introduced residual connections into the nnU-Net encoder and applied an intensified data augmentation strategy. Further modifications included experimenting with larger batch sizes and increasing the field of view. The final segmentation results were obtained by ensembling five cross-validated models.

For tumor detection, HU employed nnDetection \citep{baumgartner2021nndetection}, which automatically constructs a training pipeline based on the dataset’s characteristics. They trained the model with a batch size of 4 over 50 epochs, using down-sampled images to accommodate the large resolutions. Six resolution levels were incorporated to adjust image size, with anchor sizes scaled according to depth. Inference was performed by ensembling three models, each trained with different parameters, including variations in data augmentation strategies and anchor balance. The final ensemble predictions were merged using Weighted Box Clustering \citep{jaeger_retina_2020} with an IoU threshold of 0.2 to refine the detection output.

In the classification task, HU adopted a modified ResNet-18 \citep{he_deep_2016} architecture, specially designed to handle 3D inputs. Given the large volume of input data, they trained the model on patches of size [80, 160, 192], focusing on classifying patches as background, benign, or malignant. Training spanned 340 epochs, with a batch size of 10, and used an f1-score-based loss function to improve classification accuracy. To address the significant imbalance between foreground and background regions in the images, a 50\% foreground sampling strategy was implemented. Throughout the methodology, HU demonstrated a strong emphasis on optimizing both model architecture and data handling to maximize accuracy and efficiency across all tasks.

\subsubsection{T4: KTH Royal Institute of Technology (KTH)}
KTH developed a multi-step approach for 3D breast ultrasound analysis. For segmentation, they used the STU-NET model \citep{huang_stu-net_2023}, selecting the large model to fit within GPU constraints.

For detection, they identified the largest connected component from the segmented volume, calculating its center and dimensions along the x, y, and z axes, and assigned a score of 0.8.

For classification, KTH applied a voting-based 2.5D method. They selected the top 30 largest segmented regions, generated 2.5D images, and trained a DenseNet201 model with cross-validation. The final classification was based on voting across these images, achieving better performance compared to traditional 3D volume-based models.

\subsubsection{T5: École pour l'informatique et les techniques avancées (EPITA)}

EPITA's approach for tumor segmentation, detection, and classification centered around leveraging the 3D LowRes method within nnUNet for efficient processing of large ABUS datasets. They applied a 5-fold cross-validation to ensure robust training and mitigate overfitting. Z-score normalization was used to standardize input data, while a combination of Dice Coefficient (DC) loss and Binary Cross-Entropy (BCE) loss optimized segmentation performance.

For classification, EPITA focused on tumor volume as a key feature for distinguishing between benign and malignant tumors. They developed a model that calculated the probability of malignancy based on the tumor's size, with larger volumes increasing the likelihood of malignancy. This size-based approach provided a direct method to infer tumor malignancy.

In detection, EPITA utilized the segmentation results to generate 3D bounding boxes around the tumors. They calculated the bounding box dimensions by analyzing the mass center and the maximum and minimum coordinates of the segmented regions. Tumor presence within the bounding box was estimated by evaluating the ratio of tumor-labeled pixels to total pixels, refining detection accuracy based on the segmentation output. This integrated strategy allowed them to effectively segment, classify, and detect tumors using minimal computational resources while maintaining performance.

\subsubsection{T6: FathomX (FX)}
For segmentation, FX used an ensemble of five Swin-UNETR \citep{hatamizadeh_swin_2022} models, combining transformer-based encoders with CNN decoders. A secondary SegResNetVAE \citep{crimi_3d_2019} model ensemble was trained for small tumors. The training used a combination of dice loss, focal loss, and surface loss, with optimization through AdamW \citep{loshchilov_decoupled_2019}.

For classification, a two-step process was applied: first, a DenseNet121 \citep{huang_densely_2018} ensemble differentiated tumor-containing cubes from healthy tissue, then another ensemble classified tumors as benign or malignant.

Postprocessing involved averaging the top segmentation scores and refining masks with a conditional random field. False positives were eliminated by checking the classification confidence, and for small tumors, the secondary model predictions were used when needed. Bounding boxes were directly generated from segmentation, and cases were classified as malignant if any object scored above 90\%.

\subsubsection{T7: Philips Research (PR)}
PR employed a two-stage approach, starting with detection and classification, followed by segmentation. For detection and classification, they used the YOLOV8 architecture \citep{jocher_yolo_2023}, training three models on axial, sagittal, and coronal views. Data augmentations such as random flipping, rotation, and contrast enhancement were applied to improve robustness. A tracking algorithm assigned unique IDs to detected voxels across views, and a 3D NMS algorithm was used to refine the final detection. The built-in YOLO classifier then labeled each detection as benign or malignant, with weighted averaging applied across slices for the final malignancy score.

Building on this, PR used the MedSAM model \citep{kirillov_segment_2023, ma_segment_2024} for segmentation, fine-tuning it on the detected bounding boxes from the first stage. MedSAM, adapted for medical images, combined slice-by-slice predictions into 3D segmentation masks. Fine-tuning was conducted over 500 epochs, with the final 3D masks generated by applying the model to the bounding boxes identified in the detection stage.

\subsubsection{T8: NVIDIA (NV)}
NV utilized the Auto3DSeg framework from MONAI for automated 3D medical image segmentation and chose the SegResNet \citep{crimi_3d_2019} model for their approach. SegResNet is a U-Net-based convolutional neural network with an encoder-decoder structure and deep supervision. The model features five levels, where spatial dimensions are progressively downsized and feature sizes increased using 3x3x3 convolutions. The decoder mirrors the encoder, employing transposed convolutions for upsizing while incorporating skip connections from the encoder.

NV applied spatial augmentations such as random affine transformations, flips, intensity scaling, and noise to improve generalization. The model was optimized using Dice loss, and inference was performed using a sliding-window approach, with results resampled to the original resolution. This method allowed for efficient training while leveraging GPU resources effectively.

\subsubsection{T9: University of Girona (UDG)}

UDG developed a pipeline for 3D ABUS lesion segmentation and detection using the SAMed model \citep{zhang_customized_2023}, a version of the Segment Anything Model fine-tuned for medical images. The data was preprocessed by resizing slices to 512x512 pixels, and the model was trained using 5-fold cross-validation with geometric augmentations. Segmentation was refined by combining probability maps, selecting the top probability pixels, and expanding the largest 3D connected component. Detection was based on defining a bounding box around the lesion and calculating the mean probability within the mask.

\subsubsection{T10: Daegu Gyeongbuk Institute of Science \& Technology (DGIST)}
DGIST developed a method that utilizes multi-scale 3D crops to capture both global and local patterns for accurate detection, segmentation, and classification of abnormalities in ABUS images. Given the computational challenges of 3D CNNs, they opted for a 2D U-Net architecture \citep{navab_u-net_2015}, where input images consist of contiguous slices along the z-dimension. This approach allows the model to handle larger batch sizes, facilitating smoother optimization and focusing more effectively on malignant tissue features, despite the imbalance between foreground and background.

For segmentation, they applied both DICE and cross-entropy loss functions. While DICE loss helps balance the masking of background pixels, it can result in false positives by excluding high-probability background pixels. To address this, they included a cross-entropy loss that groups foreground and hard background pixels, allowing higher gradients to handle these difficult samples, thereby reducing false positives.

In the classification task, the model uses binary cross-entropy loss for foreground predictions, while a multi-class cross-entropy loss is applied to distinguish between background, benign, and malignant tissues. To ensure balance, benign pixels—being less common—are grouped separately in each training batch, ensuring gradient balance across all classes.

To mitigate overfitting caused by limited medical image data, DGIST employed multi-scale 3D crops. These samples, resized to uniform dimensions, allow the model to learn a mix of global and local features, enhancing boundary identification and preventing overfitting on large samples. The network architecture is based on a 2D U-Net with an encoder, bottleneck, and decoder, where skip connections help the model learn 3D patterns across slices. An additional classification branch is added to the bottleneck for identifying crops containing foreground pixels.

\subsubsection{T11: DiscerningTumor (DT)}
DT introduced a novel method for breast tumor segmentation in 3D-ABUS images by approaching the task from three different directions: x, y, and z. The first step in their approach involved converting each patient's 3D image into two-dimensional sectional images in these three planes. For training, they randomly selected images from 80 patients and used the remaining 20 as the validation set, ensuring that only images containing tumors were included in the test set. This allowed the model to focus on meaningful data, avoiding the dilution of training efficacy with too many tumor-free images. Additionally, they removed images with pixel sums below a certain threshold to eliminate irrelevant data, and they balanced the dataset by selecting an equal number of tumor and non-tumor images.

DT applied a variety of data augmentation techniques to strengthen the model's performance, including basic rotations, flips, and more advanced methods like CutOut \citep{devries_improved_2017} and CutMix \citep{yun_cutmix_2019}. These methods increased the model’s robustness by encouraging it to focus on broader features rather than relying on localized image details. The augmentations were dynamically added during the training process, further enhancing model generalization.

\begin{figure}[h]
    \centering
    \includegraphics[width=\columnwidth]{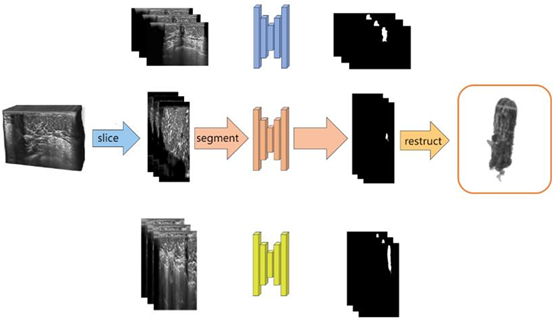} 
    \caption{Network structure diagram of T11}
    \label{fig:network_structure_t2}
\end{figure}

The team employed a multi-view network structure, shown in  Fig.  \ref{fig:network_structure_t2}, where sectional images from three directions trained three separate networks, each sharing the same architecture but with different parameters. For segmentation, DT used a modified version of the Unet++ \citep{stoyanov_unet_2018} architecture. By incorporating an attention module \citep{oktay_attention_2018} and pyramid pooling \citep{he_spatial_2015}, the network was able to suppress irrelevant regions and focus on key tumor features, while handling input size variations. This architecture allowed for better feature fusion between encoding and decoding stages, improving segmentation accuracy. The Lovász-SoftmaxLoss \citep{berman_lovasz-softmax_2018} function was used to optimize segmentation quality, especially for small tumors, by directly calculating the loss for each pixel.

For 3D reconstruction, DT combined the predictions from the three sectional planes and applied a two-step process. First, the predicted results were overlaid, and a threshold was set to differentiate tumors from the background. Then, mathematical morphology was used, involving closing and opening operations to connect tumor regions and remove noise caused by incorrect segmentation. This ensured a coherent segmentation across slices and helped to reduce errors, particularly when tumors spanned multiple slices.

\subsubsection{T12: Imperial College London (ICL)}
\begin{figure*}[ht]
    \centering
    \includegraphics[width=16cm]{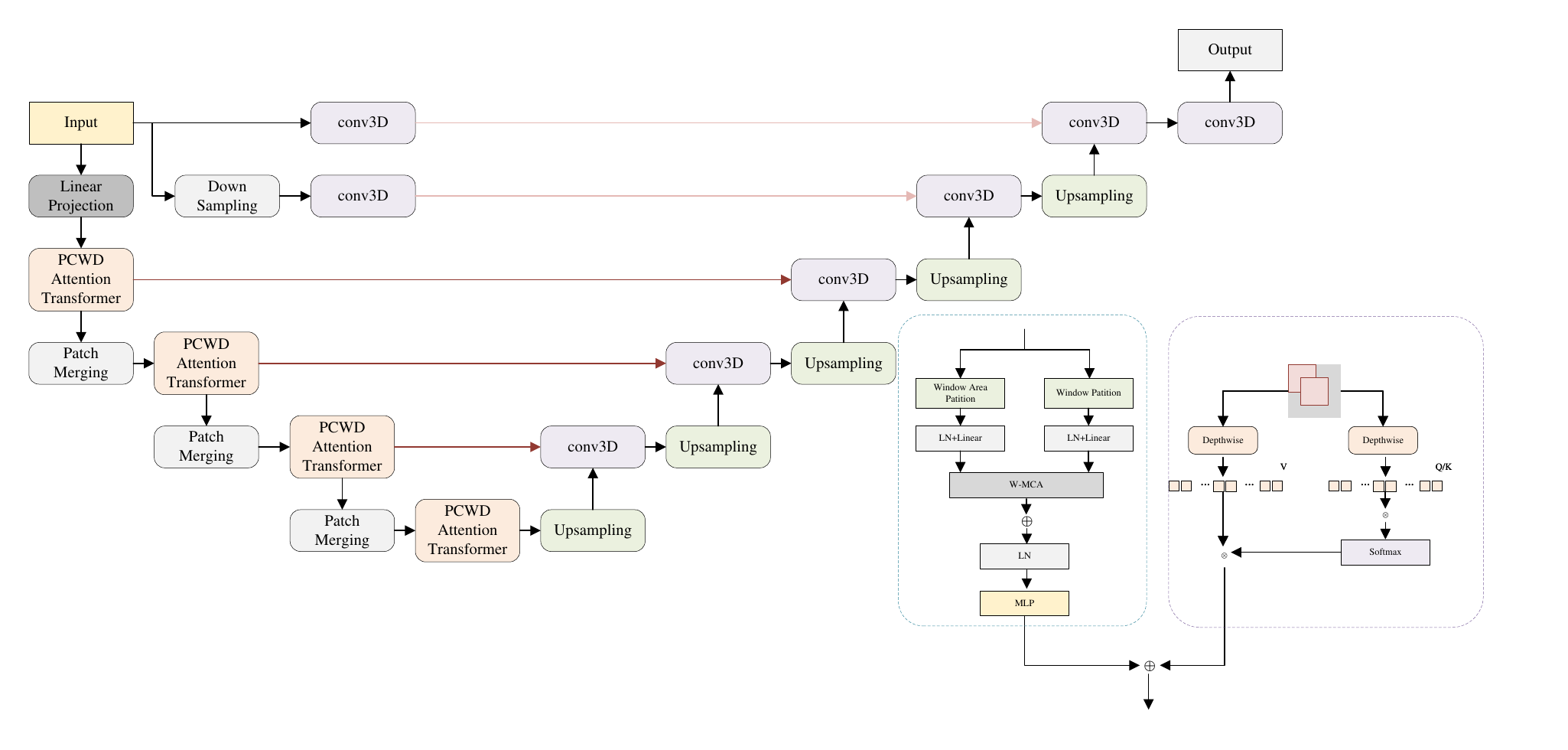} 
    \caption{The proposed model for segmentation of T12}
    \label{fig:proposed_model_t3}
\end{figure*}

ICL developed a hybrid segmentation framework that combines vision transformers with convolutional neural networks (CNNs) for efficient 3D image segmentation. Their method uses depth-wise separable convolution for spatial and channel feature extraction and incorporates a transformer-based block with cross-attention in the encoder, while the decoder uses standard 3D CNN modules. Encoder and decoder blocks are connected by concatenating feature maps, enabling better reconstruction of semantic information. They also apply deep supervision at multiple levels to enhance segmentation accuracy. The network structure of ICL is shown in  Fig. \ref{fig:proposed_model_t3}.

For training, they generated patches of size 128x128x128 and used data augmentation techniques like random cropping and Gaussian noise. The model was trained with dice loss and cross-entropy loss using the Adam optimizer. A sliding window approach was used for inference, followed by post-processing to extract the largest connected components. The model was implemented in PyTorch and trained from scratch without external pre-trained weights, using nnUNet for preprocessing, training, and validation.

\subsubsection{T13: Infervision Advanced Research Institute (IARI)}
IARI proposed a coarse-to-fine framework for breast lesion segmentation based on the ResUNet architecture. In the coarse segmentation stage, they resample the whole ABUS volume to 256x256x256 and use it as input. For fine segmentation, instead of cropping the entire region of interest (ROI) from the coarse results, they treat each connected component as a separate ROI and input them individually into the fine segmentation network. The segmented components are then merged to produce the final result.

\begin{figure}[h]
    \centering
    \includegraphics[width=\columnwidth]{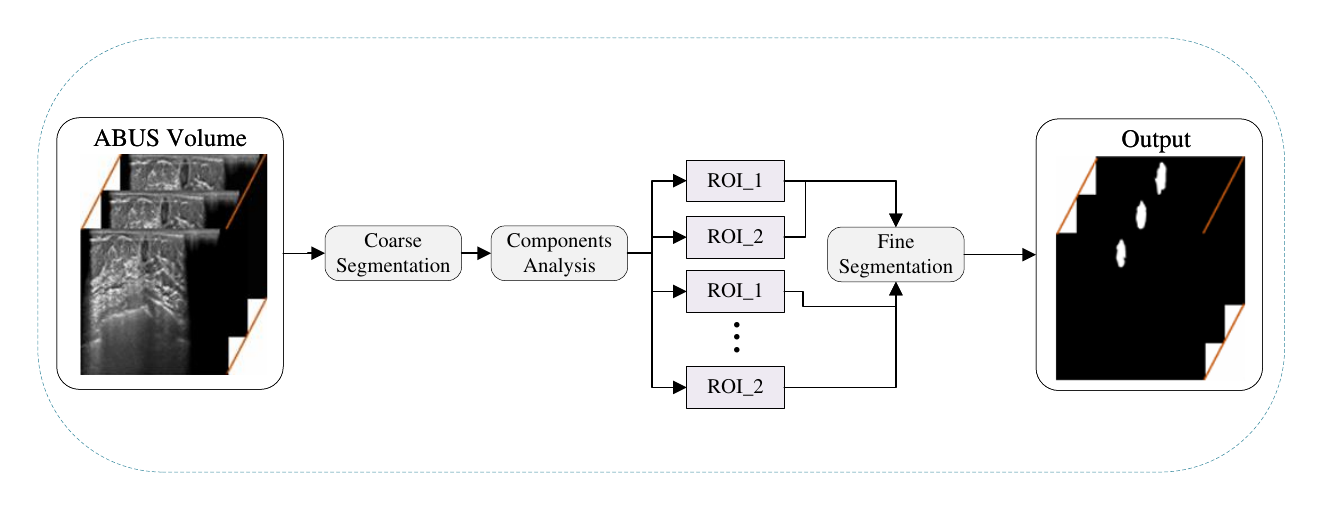}
    \caption{The framework of T13}
    \label{fig:framework_t7}
\end{figure}

Their framework addresses errors that arise during coarse segmentation by refining the results in the fine stage, minimizing interference from incorrect regions. The framework of IARI is shown in Fig. \ref{fig:framework_t7}. The network used for both stages features 4 down-sample layers, 4 up-sample layers, and an ASPP module to enhance segmentation accuracy. The loss function combines Dice loss with Cross-Entropy loss, which has proven effective in medical image segmentation tasks. This compound loss function improves robustness, using the formula:
$$
L = Dice(Pred, GT) + 0.5 \times BCE(Pred, GT)
$$
This approach aims to improve both the effectiveness and efficiency of segmentation \citep{ma_segmentation_2020}.

\subsubsection{T14: Shanghai University (SHU)}

SHU developed a system for breast lesion analysis using three specialized models. They employed VNet \citep{milletari_v-net_2016}, which leverages skip connections and a dice coefficient loss function to improve segmentation accuracy in ABUS 3D images. For classification, they utilized ResNet-101 \citep{he_deep_2016}, a deep residual network with shortcut connections that effectively classifies benign and malignant lesions. To handle detection, SHU integrated YOLOv3 \citep{redmon_yolov3_2018}, a real-time object detection system that uses multi-scale detection layers to accurately identify lesions of varying sizes. This combination of models forms an efficient and accurate solution for segmentation, classification, and detection in ABUS 3D images.

\subsubsection{T15: Macao Polytechnic University (MPU)}

MPU developed a breast tumor detection framework called NoduleNet, consisting of feature extraction, multi-region proposal, and false positive reduction. The feature extraction uses six 3D Res Block filters modeled after ResNet, applying convolutions, normalizations, activations, and max pooling. The multi-region proposal stage employs a Region Proposal Network (RPN) to generate candidate tumor regions using 3D convolutions, classifying foreground and background, and regressing anchor parameters for different-sized cubes. Non-Maximum Suppression (NMS) is applied to reduce false positives, and a Region of Interest (ROI) function extracts feature maps for further classification and regression through an RCNN network. NMS is applied again, with losses minimized to improve classification and regression accuracy.

To evaluate the performance, the Free-response ROC (FROC) curve was applied. On the validation set, the framework achieved maximum recall and average recall rates of 88.89\% and 86\%, respectively. On the testing set, the rates were 80\% and 74.29\%. Ablation experiments demonstrated that using the RPN network improved the average recall rate by 3.8\%, and the false positives reduction framework further increased the rate by 0.028\%.

\subsubsection{T16: University of Amsterdam (UA)}

UA developed a framework for breast tumor segmentation and detection using a combination of 2D and 3D techniques. For segmentation, they employed a 2D approach to handle the large 3D ABUS volumes, converting them into 2D slices along the Z-axis. This approach, combined with the UNet architecture \citep{navab_u-net_2015}, efficiently processed the data with lower computational demands than 3D convolution networks. The UNet model applied convolutions, ReLU activations, max pooling, and upsampling, with dropout added to prevent overfitting. Postprocessing involved test-time augmentation (TTA) and 3D connected component analysis (3DCCA) to refine segmentation results. TTA used flip averaging to augment test images and merge predictions, while 3DCCA identified the largest connected tumor group for the final segmentation.

For detection, UA used a YOLOv5 architecture enhanced with ghost convolutions. Ghost convolutions \citep{han_ghostnet_2020} replaced traditional convolutions to generate more features with fewer parameters, improving model efficiency. The detection process involved running inference on 2D slices along the Z-axis, selecting the largest group of consecutive slices to determine tumor boundaries, and calculating the tumor probability based on the mean confidence of these slices. This approach provided accurate detection and efficient segmentation for ABUS images.

\subsubsection{T17: University of California, Los Angeles (UCLA)}

UCLA used a 3D U-Net architecture for breast tumor segmentation in 3D ABUS volumes. The data was whitened and split into 64x64x64 patches due to memory constraints, with edge patches excluded. The model was trained using a combination of Dice loss and Binary Cross Entropy loss, with a learning rate of 0.001 and a batch size of 16. After training, the best model weights were used to reconstruct the predicted patches back into full volumes for testing.

\subsubsection{T18: Xi’an Jiaotong-Liverpool University (XJTLU)}

XJTLU used a combination of MDA-net, Unet, nnU-Net, and DAF3D models for breast tumor segmentation, employing both 2.5D (three consecutive slices) and 3D (entire volumes) approaches. Unet extracts features and restores resolution with upsampling and skip connections. DAF3D refines features using a 3D attention mechanism to handle challenges in ultrasound images. nnU-Net automatically adjusts its training process based on the dataset, providing an adaptable framework for medical image segmentation.

\subsection{Evaluation Results and Ranking Analysis}

To evaluate the segmentation, classification, and detection tasks, we employed a series of metrics designed to capture the unique performance aspects of each task. The primary metrics used include the Dice coefficient (DICE) and Hausdorff distance (HD) for segmentation, accuracy (ACC) and area under the curve (AUC) for classification, and the Free-Response Receiver Operating Characteristic (FROC) for detection.

\subsubsection*{Scoring Formulas for Each Task}
- \textbf{Segmentation Task}: The segmentation task score is calculated using DICE and HD metrics. First, we exclude teams without valid results. The remaining scores are then normalized using min-max normalization:
  \[
  \text{Norm}(x) = \frac{x - \text{min}(x)}{\text{max}(x) - \text{min}(x)}
  \]
  The final score for segmentation is calculated as:
  \[
  \text{Segmentation Score} = \frac{1 + \text{Norm\_DICE} - \text{Norm\_HD}}{2}
  \]
  Table~\ref{tab:segmentation_scores} shows the scores for segmentation.

- \textbf{Classification Task}: For classification, we employ ACC and AUC metrics. Teams with invalid results are removed, and remaining scores undergo min-max normalization. The classification score is computed as:
  \[
  \text{Classification Score} = \frac{\text{Norm\_ACC} + \text{Norm\_AUC}}{2}
  \]
  Table~\ref{tab:classification_scores} displays the classification scores.

- \textbf{Detection Task}: Detection performance is evaluated using the FROC metric, focusing on average sensitivity at various false positive (FP) levels. The final detection score is computed by averaging the sensitivity across FP rates of 0.125, 0.25, 0.5, 1, 2, 4, and 8. These values are then normalized using min-max normalization. The detection scores are shown in Table~\ref{tab:detection_scores}.

Some teams received an 'inf' value in HD, which complicates fair evaluation. To address this, we created a secondary leaderboard by replacing 'inf' scores with the worst HD score from valid results, multiplied by 105\%. This allows a comprehensive ranking while minimizing unfair penalization. The adjusted segmentation scores are shown in Table~\ref{tab:segmentation_penalty_scores}.

The overall performance considers only teams that have submitted valid results for all metrics. The final overall result is computed as:

\begin{align}
\text{Overall Score} = & \ \frac{1 + \text{Norm\_DICE} - \text{Norm\_HD}}{2} \nonumber \\
& + \frac{\text{Norm\_ACC} + \text{Norm\_AUC}}{2} \nonumber \\
& + \text{Norm\_FROC}
\end{align}

For the Segmentation task, we have substituted any `inf` values with the worst HD score among all valid results, multiplied by 105\%. This adjustment enables fair ranking of teams with `inf` results. Table~\ref{tab:overall_scores} presents the final overall rankings for all teams.

\begin{table*}[h!]
\centering
\caption{Segmentation task scores for the 10 qualified teams, including DICE (Dice Similarity Coefficient), Norm\_DICE (Normalized Dice Similarity Coefficient), HD (Hausdorff Distance), Norm\_HD (Normalized Hausdorff Distance), and Seg\_Score (Segmentation Score).}
\label{tab:segmentation_scores}
\begin{tabular}{lccccc}
\toprule
Team       & DICE   & Norm\_DICE & HD       & Norm\_HD & Seg\_Score \\
\midrule
T2 & 0.6147 & 1.0000     & 90.5339  & 0.0557   & 0.9722     \\
T9 & 0.5853 & 0.9050     & 80.1817  & 0        & 0.9525     \\
T5 & 0.5377 & 0.7509     & 96.5050  & 0.0878   & 0.8316     \\
T18 & 0.4890 & 0.5933    & 81.7367  & 0.0084   & 0.7925     \\
T6 & 0.5400 & 0.7583     & 121.1640 & 0.2204   & 0.7689     \\
T3 & 0.5616 & 0.8283    & 162.9371 & 0.4451   & 0.6916     \\
T16 & 0.4412 & 0.4386    & 101.4036 & 0.1141   & 0.6622     \\
T7 & 0.4981 & 0.6227     & 153.0743 & 0.3920   & 0.6154     \\
T12 & 0.4665 & 0.5204    & 266.1207 & 1        & 0.2602     \\
T13 & 0.3057 & 0.0000    & 203.4005 & 0.6627   & 0.1687     \\
\bottomrule
\end{tabular}
\end{table*}

\begin{table*}[h!]
\centering
\caption{Segmentation Task Scores with Penalty Adjustment: HD scores marked as "inf" were replaced by the worst HD score from all valid results for each case, multiplied by 105\%.}
\label{tab:segmentation_penalty_scores}
\begin{tabular}{lccccc}
\toprule
Team       & DICE   & Norm\_DICE & HD       & Norm\_HD & Seg\_Score \\
\midrule
T8  & 0.6020 & 0.9590     & 82.8654  & 0.0144   & 0.9723     \\
T2  & 0.6147 & 1.0000     & 90.5339  & 0.0557   & 0.9722     \\
T9  & 0.5853 & 0.9050     & 80.1817  & 0        & 0.9525     \\
T1  & 0.5861 & 0.9075     & 117.1939 & 0.1991   & 0.8542     \\
T5  & 0.5377 & 0.7509     & 96.5050  & 0.0878   & 0.8316     \\
T10 & 0.5342 & 0.7395     & 105.0751 & 0.1339   & 0.8028     \\
T18 & 0.4890 & 0.5933     & 81.7367  & 0.0084   & 0.7925     \\
T6  & 0.5400 & 0.7583     & 121.1640 & 0.2204   & 0.7689     \\
T4  & 0.5890 & 0.9169     & 159.0311 & 0.4241   & 0.7464     \\
T3 & 0.5616 & 0.8283     & 162.9371 & 0.4451   & 0.6916     \\
T16 & 0.4412 & 0.4386     & 101.4036 & 0.1141   & 0.6622     \\
T7  & 0.4981 & 0.6227     & 153.0743 & 0.3920   & 0.6154     \\
T12 & 0.4665 & 0.5204     & 266.1207 & 1        & 0.2602     \\
T13 & 0.3057 & 0.0000     & 203.4005 & 0.6627   & 0.1687     \\
\bottomrule
\end{tabular}
\end{table*}

\begin{table*}[h!]
\centering
\caption{Classification task scores for the 8 qualified teams, including ACC (Accuracy), Norm\_ACC (Normalized Accuracy), AUC (Area Under the Curve), Norm\_AUC (Normalized Area Under the Curve), and Cls\_Score (Classification Score).}
\label{tab:classification_scores}
\begin{tabular}{lccccc}
\toprule
Team       & ACC   & Norm\_ACC & AUC       & Norm\_AUC & Cls\_Score \\
\midrule
T1      & 0.7571 & 1.0000     & 0.8892  & 1.0000   & 1.0000     \\
T4      & 0.7429 & 0.9333     & 0.7708  & 0.6321   & 0.7827     \\
T3     & 0.7286 & 0.8667     & 0.7733  & 0.6399   & 0.7533     \\
T10        & 0.7143 & 0.8000     & 0.7642  & 0.6114   & 0.7057     \\
T2           & 0.6429 & 0.4667     & 0.6558  & 0.2746   & 0.3706     \\
T7       & 0.6000 & 0.2667     & 0.6425  & 0.2332   & 0.2499     \\
T5     & 0.5429 & 0.0000     & 0.5775  & 0.0311   & 0.0155     \\
T6      & 0.5429 & 0.0000     & 0.5675  & 0.0000   & 0.0000     \\
\bottomrule
\end{tabular}
\end{table*}

\begin{table*}[h!]
\centering
\caption{Detection task scores for the 10 qualified teams, including FROC (Free-response Receiver Operating Characteristic) and Det\_Score (Detection Score).}
\label{tab:detection_scores}
\begin{tabular}{lcc}
\toprule
Team       & FROC   & Det\_Score \\
\midrule
T1  & 0.8468 & 1.0000     \\
T3 & 0.7704 & 0.8323     \\
T2  & 0.7303 & 0.7442     \\
T9  & 0.6459 & 0.5589     \\
T7  & 0.6441 & 0.5550     \\
T5  & 0.6383 & 0.5423     \\
T6  & 0.6153 & 0.4918     \\
T4  & 0.6067 & 0.4729     \\
T15 & 0.5327 & 0.3104     \\
T16 & 0.3913 & 0.0000     \\
\bottomrule
\end{tabular}
\end{table*}

\begin{table*}[h!]
\centering
\caption{Overall result of the quantitative evaluation for the 7 qualified teams, including DICE (Dice Similarity Coefficient), Norm\_DICE (Normalized Dice Similarity Coefficient), HD (Hausdorff Distance), Norm\_HD (Normalized Hausdorff Distance), ACC (Accuracy), Norm\_ACC (Normalized Accuracy), AUC (Area Under the Curve), Norm\_AUC (Normalized Area Under the Curve), FROC (Free-response Receiver Operating Characteristic), and Norm\_FROC (Normalized FROC).}
\label{tab:overall_scores}
\resizebox{\textwidth}{!}{%
\begin{tabular}{lccccccccccc}
\toprule
Team       & DICE   & Norm\_DICE & HD       & Norm\_HD & ACC & Norm\_ACC & AUC & Norm\_AUC & FROC & Norm\_FROC & Overall \\
\midrule
T1  & 0.5861 & 0.7547     & 117.1939  & 0.3682   & 0.7571 & 1.0000 & 0.8892 & 1.0000 & 0.8468 & 1.0000 & 2.6932     \\
T2  & 0.6147 & 1.0000     & 90.5339   & 0.0000   & 0.6429 & 0.4667 & 0.6558 & 0.2746 & 0.7303 & 0.5148 & 1.8854     \\
T3 & 0.5616 & 0.5449     & 162.9371  & 1.0000   & 0.7286 & 0.8667 & 0.7733 & 0.6399 & 0.7704 & 0.6818 & 1.7075     \\
T4  & 0.5890 & 0.7796     & 159.0311  & 0.9461   & 0.7429 & 0.9333 & 0.7708 & 0.6321 & 0.6067 & 0.0000 & 1.1995     \\
T5  & 0.5377 & 0.3397     & 96.5050   & 0.0825   & 0.5429 & 0.0000 & 0.5775 & 0.0311 & 0.6383 & 0.1316 & 0.7758     \\
T6  & 0.5400 & 0.3593     & 121.1640  & 0.4230   & 0.5429 & 0.0000 & 0.5675 & 0.0000 & 0.6153 & 0.0358 & 0.5039     \\
T7  & 0.4981 & 0.0000     & 153.0743  & 0.8638   & 0.6000 & 0.2667 & 0.6425 & 0.2332 & 0.6441 & 0.1558 & 0.4738     \\
\bottomrule
\end{tabular}%
}
\end{table*}

\begin{figure*}[h!]
    \centering
    \begin{subfigure}{0.49\linewidth}
        \centering
        \includegraphics[width=\textwidth]{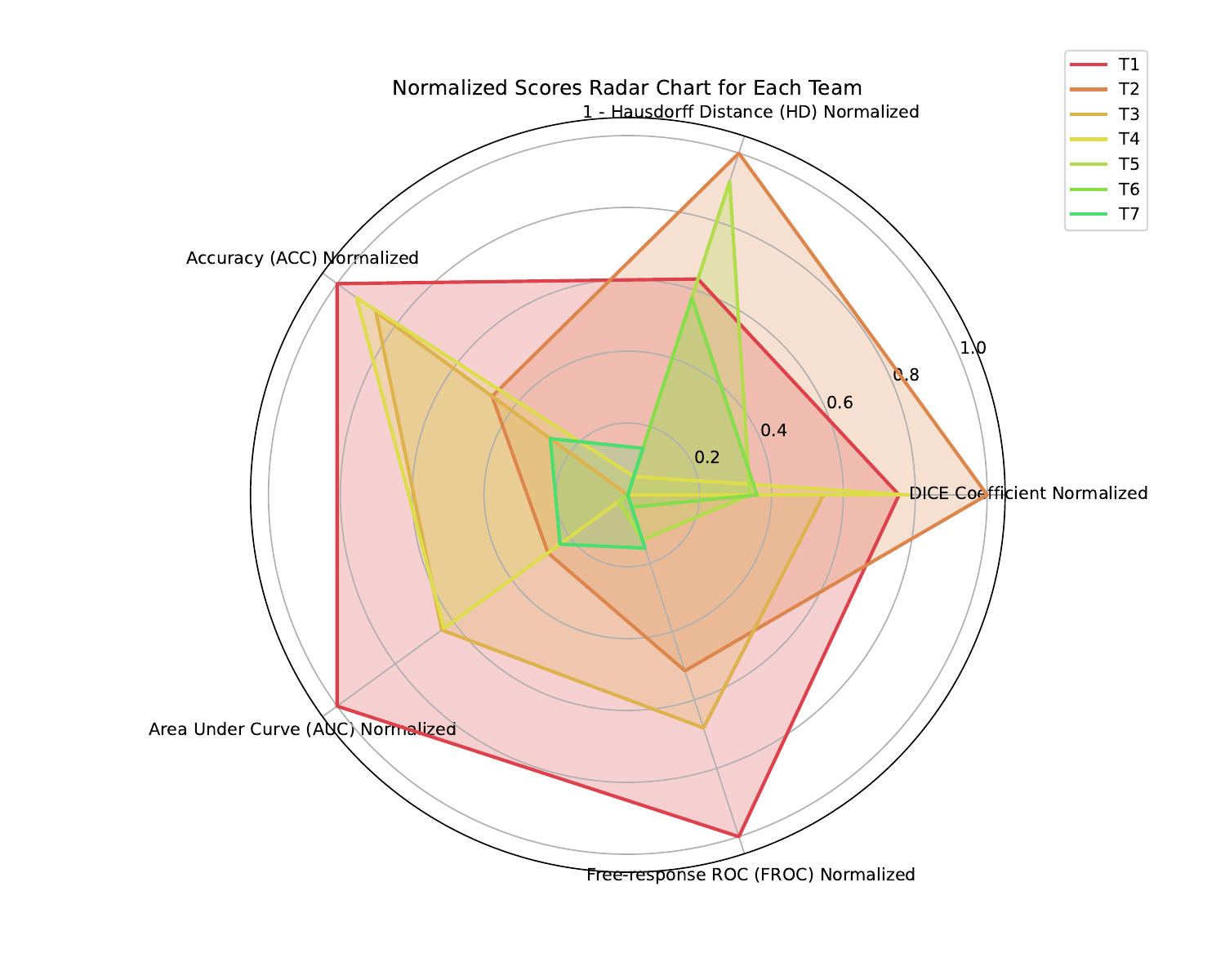} 
        \caption{Radar chart displaying the performance across five metrics: DICE, Norm DICE, HD, Norm HD, and FROC. Note that for HD, which benefits from lower values, the metric used is transformed to $1 - \text{HD}$.}
    \end{subfigure}
    \hfill
    \begin{subfigure}{0.49\linewidth}
        \centering
        \includegraphics[width=\textwidth]{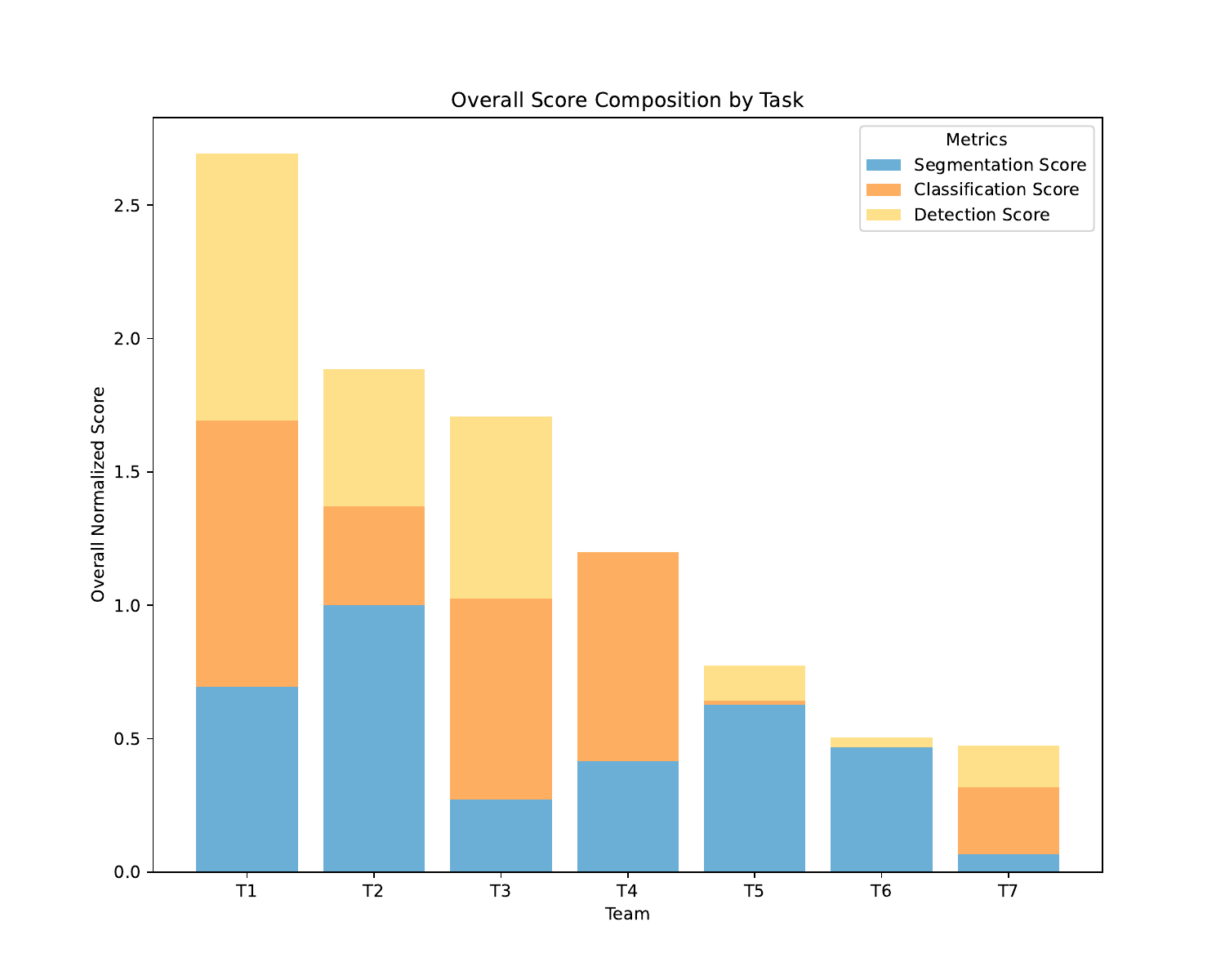} 
        \caption{Stacked bar chart illustrating the overall score and the contributions from segmentation, classification, and detection tasks.}
    \end{subfigure}
    \caption{Figures illustrating the final scores of the teams that successfully completed all three tasks. (a) Radar chart showing individual performance across five metrics; (b) Stacked bar chart comparing the overall score and task-specific contributions for each team.}
    \label{fig:overall_performance}
\end{figure*}

To provide a clearer view of the final scores achieved by the teams, we present two figures, Fig.~\ref{fig:overall_performance} (a) and (b), showcasing the results of the seven teams that successfully completed all three tasks. Figure (a) is a radar chart that displays the performance across five metrics: Norm ACC, Norm DICE, Norm AUC, Norm HD, and FROC. It is important to note that for the HD metric, which benefits from lower values, we have transformed it using \(1 - \text{HD}\). 

Figure (b) illustrates a stacked bar chart that compares the overall scores and the contributions from the three different tasks. It is evident that Team T1 excels in both Classification and Detection tasks while showing a slight deficiency in Segmentation. However, overall, T1 demonstrates the most balanced and robust performance across all metrics.

\subsubsection{Segmentation Task}

\begin{figure*}[h!]
    \centering
    \begin{subfigure}{0.49\linewidth}
        \centering
        \includegraphics[width=\textwidth]{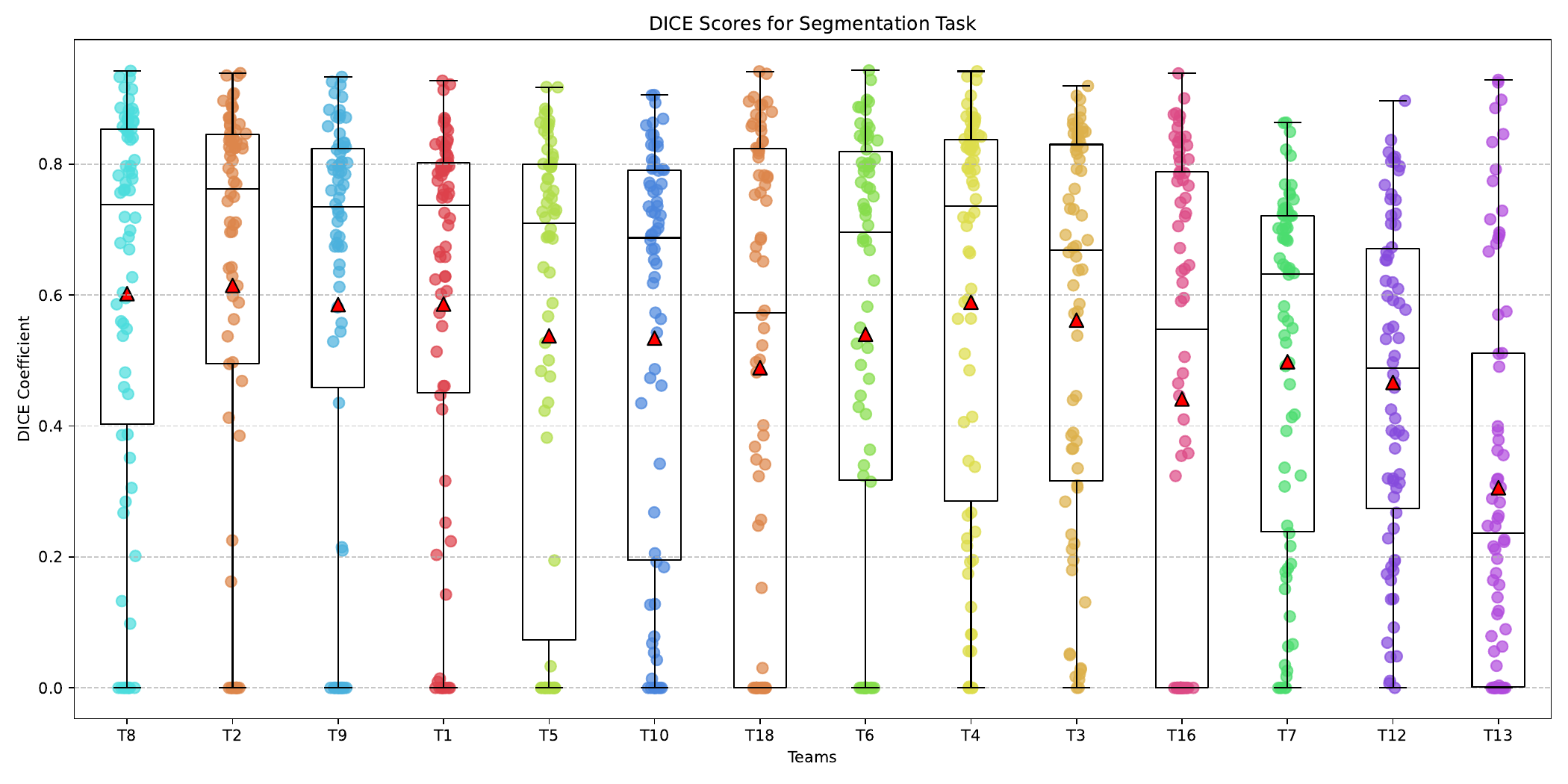}
        \caption{DICE Box Plot: Team performance per case. The distribution, median, and mean values are shown.}
    \end{subfigure}
    \hfill
    \begin{subfigure}{0.49\linewidth}
        \centering
        \includegraphics[width=\textwidth]{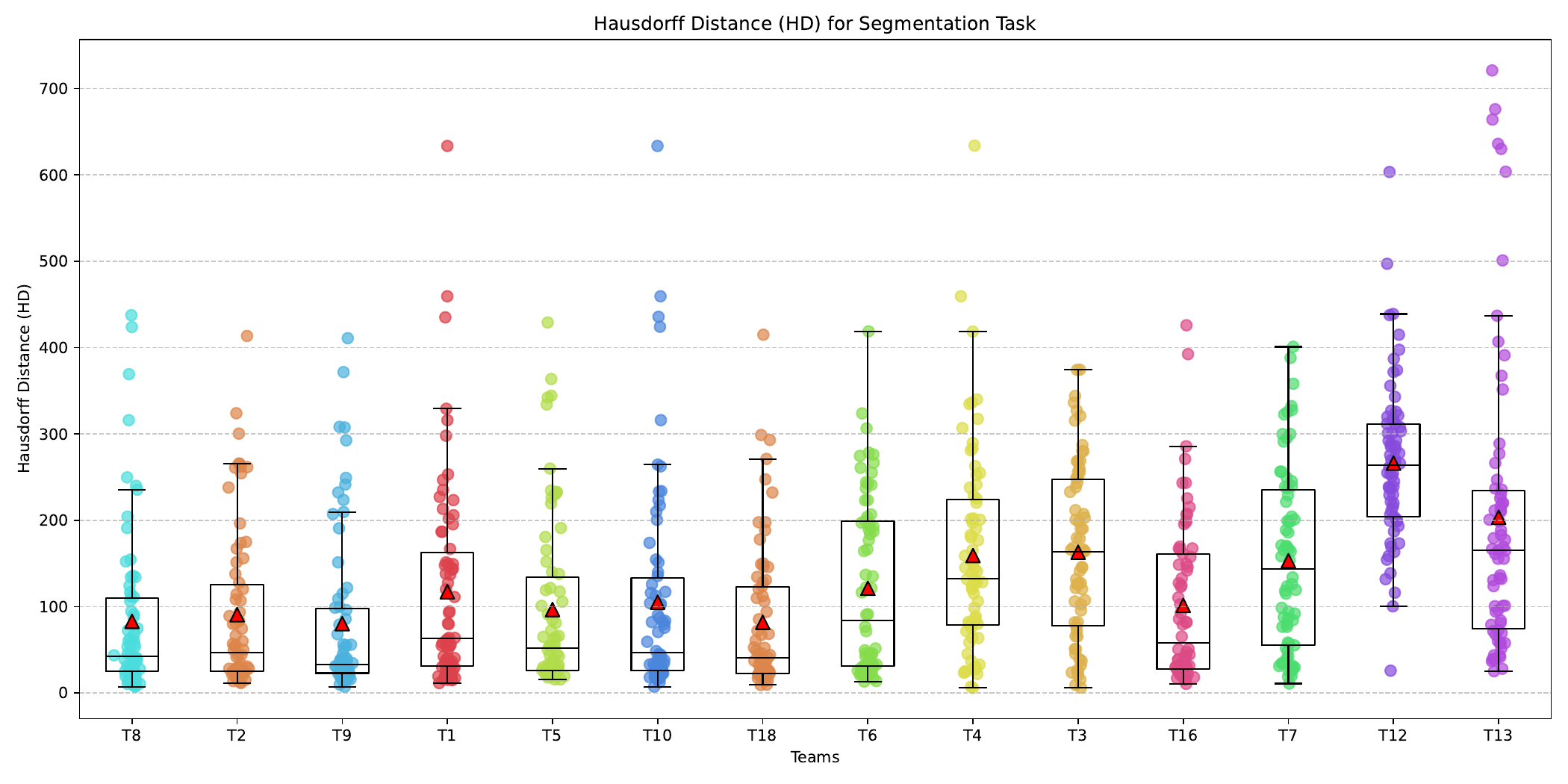}
        \caption{HD Box Plot: Team performance per case. The distribution and median values are shown. HD is penalized for higher values.}
    \end{subfigure}
    \vspace{0.5cm} 
    \begin{subfigure}{0.98\linewidth}
        \centering
        \includegraphics[width=\textwidth]{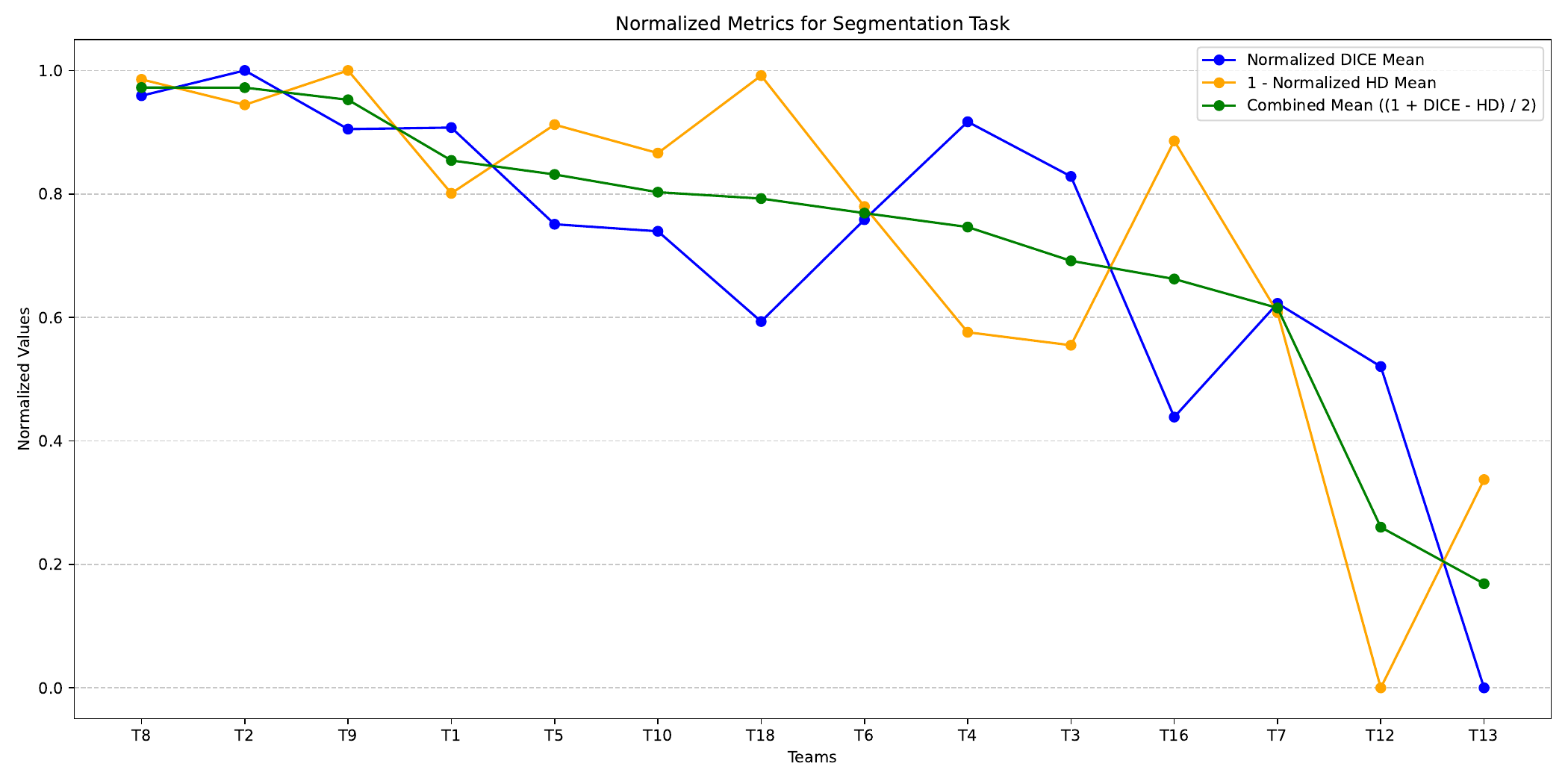}
        \caption{Combined Line Plot: Blue line is Norm\_DICE, orange line is 1 - Norm\_HD, and green line is the final score ((1 + Norm\_DICE - Norm\_HD) / 2).}
    \end{subfigure}
    \caption{Visualization of segmentation task performance. Subfigures show (a) the DICE box plot, (b) the HD box plot, and (c) the line plot summarizing team scores. The x-axis represents the teams, and the y-axis shows the values for DICE, HD, and final scores.}
    \label{fig:figure_seg}
\end{figure*}

As shown in Table \ref{tab:segmentation_penalty_scores} and Figure \ref{fig:figure_seg} (a-b), each team’s performance in the segmentation task was evaluated based on their DICE and HD scores. DICE, a standard metric for segmentation accuracy, measured the overlap between predicted and ground truth regions, while HD assessed the boundary accuracy. For teams with infinite HD values, we applied a penalization strategy by substituting the “inf” value with 105\% of the worst valid HD score to ensure fair comparison.

It is noted that in the segmentation results without fixed penalization for Inf HD, shown in Table \ref{tab:segmentation_scores}, T2  ranks first. While in Table \ref{tab:segmentation_penalty_scores}, teams T8, T2, and T9 achieved the top three normalized DICE scores, indicating high overlap accuracy in their segmentation results. However, T18  and T7, while excelling in DICE, showed higher variability in their HD scores, suggesting less consistent boundary precision. Notably, T9 led in HD performance, with the lowest HD score, suggesting a strong emphasis on boundary accuracy in its approach.

The statistical distribution shown in Figure \ref{fig:figure_seg} illustrates considerable variation in teams' DICE and HD scores for the segmentation task. Top-performing teams were able to achieve a balance between high overlap and boundary precision. T8, while not having the lowest HD or highest DICE, demonstrated balanced performance across both DICE and HD metrics, indicating a well-rounded approach to segmentation.

These results reveal that a high DICE score does not necessarily correlate with a low HD score, underscoring the importance of evaluating both metrics for a comprehensive assessment of segmentation quality. The use of both DICE and HD, particularly with penalization for infinite HD scores, enables a more balanced and thorough evaluation of each team’s segmentation capability.

\subsubsection{Classification Task}

\begin{figure*}[h!]
    \centering
    \includegraphics[width=0.9\textwidth]{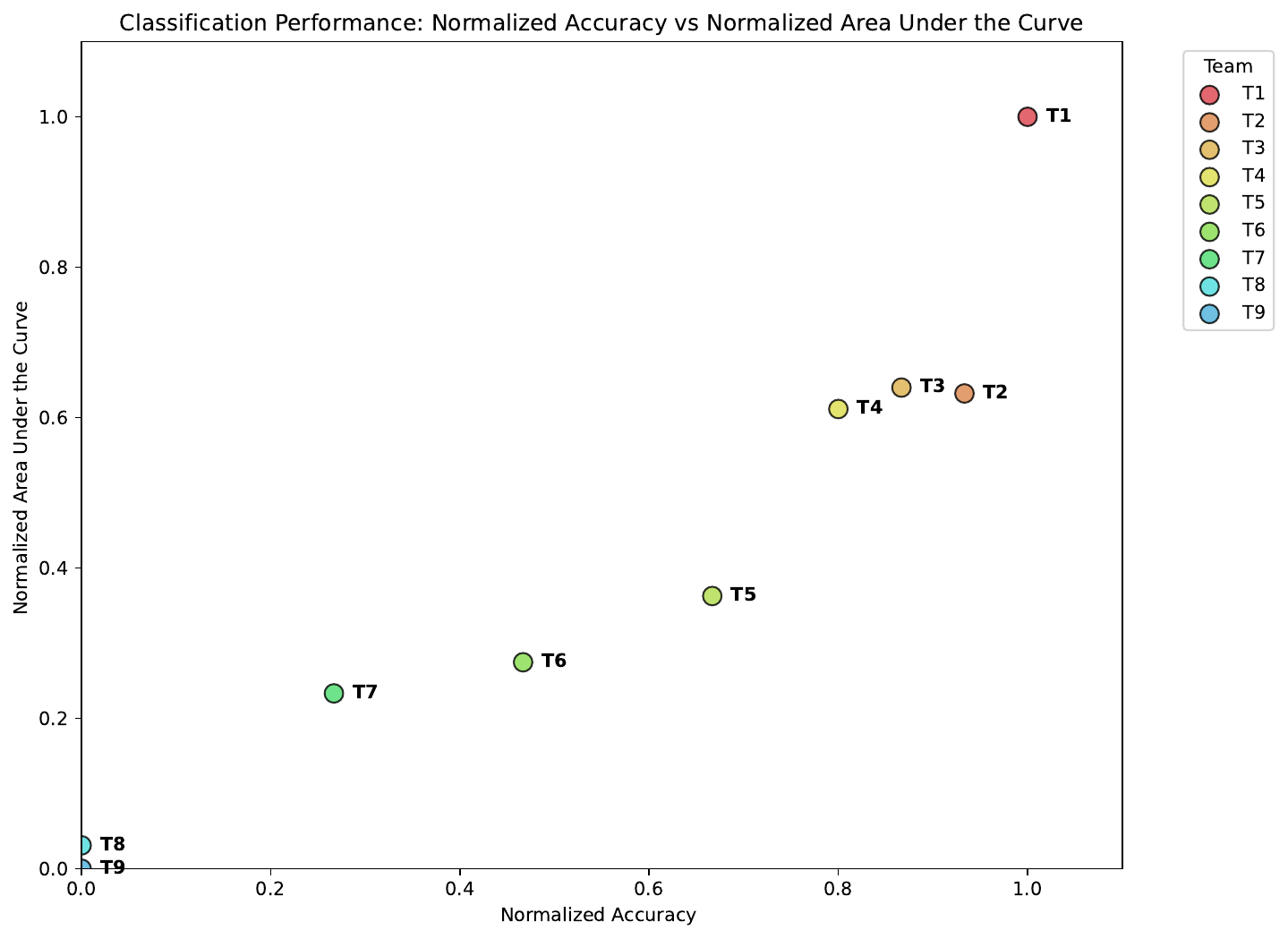}
    \caption{Classification task scatter plot, with Accuracy on the x-axis and Area Under the Curve on the y-axis. Each point represents a team.}
    \label{fig:classification_scatterplot}
\end{figure*}

As shown in Table \ref{tab:classification_scores} and Figure \ref{fig:classification_scatterplot}, each team's classification performance was evaluated based on their Accuracy (ACC) and Area Under the Curve (AUC) scores. ACC measures the overall classification correctness, while AUC reflects the model's ability to distinguish between classes.

From Table \ref{tab:classification_scores}, we observe that T1 achieved the highest normalized scores for both ACC and AUC, indicating an excellent balance between classification correctness and class separation. Following closely, T4 and T3 ranked in the top three positions, showing strong classification performance across both metrics. 

In contrast, T6 and T5 exhibited low normalized scores for both ACC and AUC, indicating challenges in achieving accurate classification results. The scatter plot in Figure \ref{fig:classification_scatterplot} further illustrates the distribution of each team’s classification scores, with Team Shiontao clearly leading, particularly in AUC.

\subsubsection{Detection Task}
As shown in Table \ref{tab:detection_scores}, each team’s detection performance was evaluated using FROC. The Free-Response Receiver Operating Characteristic (FROC) assesses detection sensitivity at various false positive rates, providing a comprehensive view of a model’s capability to detect true positives across a range of thresholds. 

T1, T3 and T2 attained the highest FROC values, indicating strong sensitivity in detecting relevant cases while managing false positives effectively. 

The line graph in Figure 11 presents the FROC performance of different teams across varying false positive rates, offering a comprehensive comparison of sensitivity and false positive trade-offs. Among the teams, T1 consistently demonstrates superior performance, achieving the highest average recall of 0.8468, showcasing its robustness in balancing high sensitivity with low false positives.

T3 and T2 follow closely with average recalls of 0.7704 and 0.7303, respectively, indicating strong detection capabilities at various thresholds. T9 and T7 exhibit competitive performance, with average recalls of 0.6459 and 0.6441, suggesting a moderate balance between sensitivity and precision. Conversely, teams such as T16 and T15, with average recalls of 0.3913 and 0.5327, respectively, show room for improvement, especially in scenarios with higher false positive rates.

A key observation from the figure is the variation in the starting points and slopes of the FROC curves. Teams like T1 and T3 have curves that rapidly rise, indicating strong recall rates even at very low false positive thresholds. In contrast, teams like T16 show slower improvement, with flatter curves that indicate lower sensitivity overall.

\begin{figure*}[h!]
    \centering
    \includegraphics[width=0.9\textwidth]{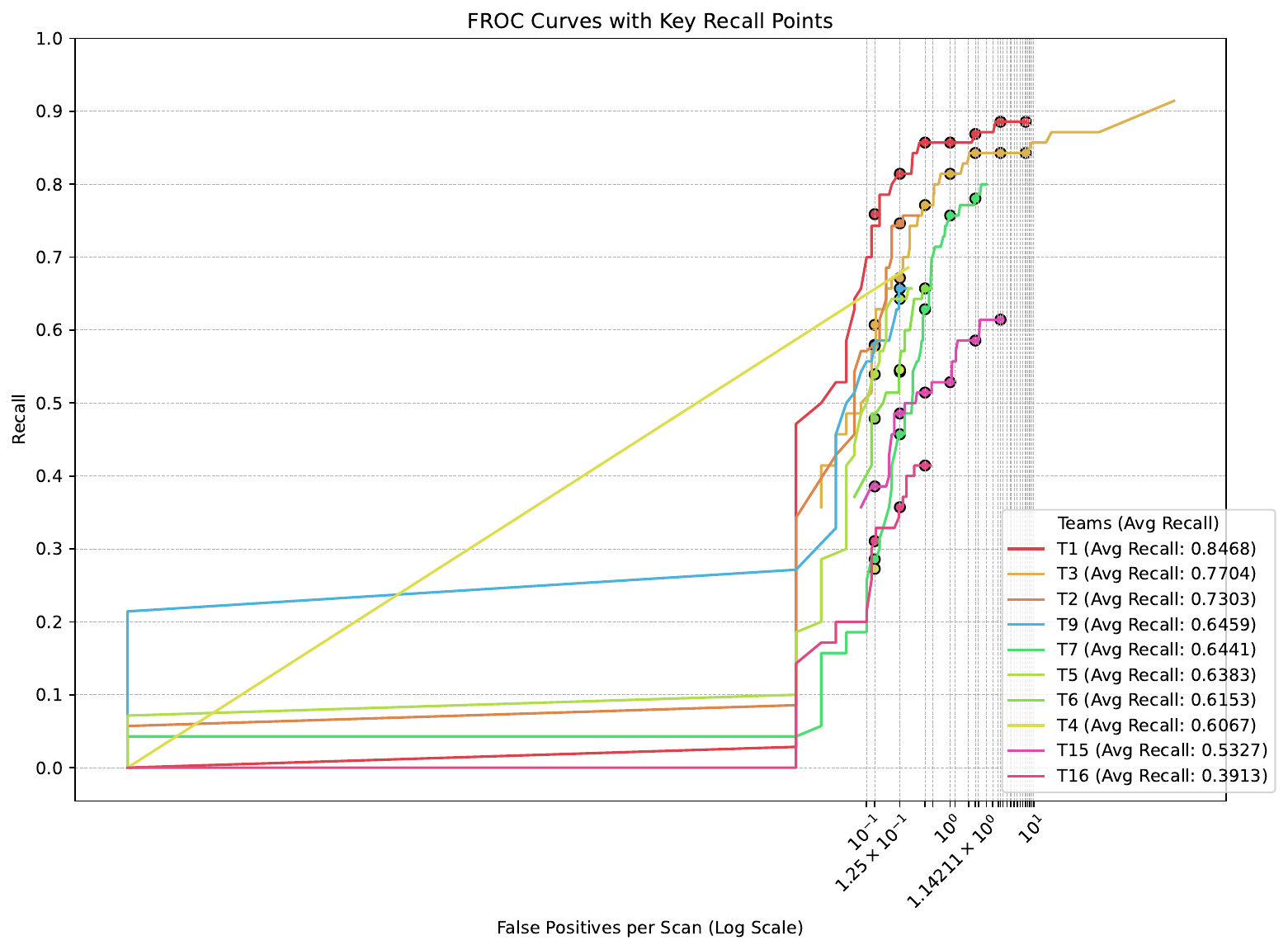}
    \caption{FROC curve for the detection task, displaying sensitivity (recall) on the y-axis and False Positives Per Image (FPPI) on the x-axis. The x-axis is plotted on a logarithmic scale to better visualize the performance across a wide range of false positive rates. Each curve represents a team’s sensitivity as a function of false positives, where higher and steeper curves indicate better overall performance. Markers on the curves denote key recall points at specific FPPI thresholds (e.g., 0.125, 0.25, 0.5, 1, etc.), providing additional insight into team performance at these critical thresholds. Teams with consistently high sensitivity, as indicated by higher average recall values, demonstrate superior detection capabilities. The legend is sorted by average recall values in descending order, with top-performing teams listed first, reflecting their robustness in achieving high recall while minimizing false positives.}
    \label{fig:froc_curve}
\end{figure*}

\section{Discussion}

\subsection{Handling 'inf' Results in HD Metrics}
In some cases, teams encountered an 'inf' result in the HD metric, which posed a significant challenge for fair comparison. Addressing these 'inf' results required a balanced approach to ensure fairness while accurately reflecting team performance.

A common strategy is to penalize 'inf' results by assigning a fixed value. However, this approach introduces subjectivity in selecting the penalty value and risks unfairly disadvantaging teams with robust results. On the other hand, outright exclusion of teams with 'inf' results overlooks their overall performance and creates an incomplete leaderboard.

To address these concerns, we adopted a method that maintains fairness and integrity in the rankings. Specifically, for cases with 'inf' results, we replaced the 'inf' value with the worst valid HD score from all teams for that case, multiplied by 105\%. This approach ensures that the penalized score is slightly worse than the poorest valid result, appropriately reflecting the challenges faced in those cases without excessively skewing the results.

\subsection{Improving Strategies for Segmentation}

Segmentation of 3D ABUS data in the TDSC-ABUS 2023 Challenge presented challenges such as class imbalance, small lesion detection, and computational constraints. To overcome these issues, participants implemented diverse and innovative strategies, which are summarized below.

\textbf{Model Customization and Architectural Enhancements.}  
Most teams utilized well-established architectures like UNet or nnU-Net as their segmentation backbones, but many introduced modifications to adapt to the specific characteristics of ABUS data. Common enhancements included adding residual connections to improve feature propagation and prevent vanishing gradients, and employing ensemble models to increase robustness across diverse lesion types. Some approaches incorporated multi-scale designs, where 3D cropping techniques captured both global and local patterns, leading to more precise boundary delineation and better handling of lesions of varying sizes.

\textbf{Data Augmentation.}  
Data augmentation played a critical role in improving model generalization and robustness. Techniques such as flipping, rotation, intensity scaling, and noise injection were commonly applied to increase variability in the training data. Additionally, some strategies focused on refining segmentation by leveraging probability maps, where only the most confident predictions were retained, combined with connected component analysis to ensure coherent and accurate segmentation.

\textbf{Loss Functions and Optimization.}  
To address the inherent class imbalance in ABUS datasets, many teams adopted sophisticated loss functions. A common approach was combining dice loss and cross-entropy loss to balance segmentation of background and lesion regions effectively. Others incorporated focal loss to penalize hard-to-classify samples more strongly, improving the model's focus on challenging regions. Some strategies also prioritized sampling of foreground regions during training, ensuring that lesions received adequate attention despite their smaller representation in the data.

\textbf{Efficient Handling of Large 3D ABUS Data.}  
The high resolution of 3D ABUS data posed computational challenges, prompting several teams to adopt innovative methods to manage memory constraints. Sliding window inference was frequently employed to process large volumes without compromising on resolution. Additionally, some strategies involved using low-resolution processing pipelines or multi-stage inference workflows, where coarse segmentation was refined in subsequent steps to achieve both efficiency and accuracy.

\textbf{Integration with Other Tasks.}  
In many cases, segmentation was integrated with detection and classification tasks to create a comprehensive analysis pipeline. For example, segmentation results were often used as a precursor to detection, guiding the generation of bounding boxes or improving lesion localization. This integration helped enhance the overall accuracy and robustness of the system, particularly when combined with multi-task learning frameworks.

\textbf{Ensemble and Post-Processing.}  
Post-processing was a critical step in refining segmentation results. Ensembles of models trained with different initialization or data splits were commonly used to achieve more consistent predictions. Additionally, techniques like conditional random fields or clustering algorithms were applied to refine segmentation masks, remove false positives, and ensure the outputs were both accurate and reliable.

These strategies showcase the variety and innovation in tackling the challenges of 3D ABUS data segmentation. By integrating advanced architectural designs, tailored loss functions, and efficient data management techniques, participants made significant strides in solving complex segmentation problems. Their efforts contribute to improving the accuracy and reliability of breast cancer diagnosis through ultrasound imaging.

\subsection{Improving Strategies for Classification}

Classification in 3D breast ultrasound data focused on distinguishing benign and malignant lesions, a task complicated by class imbalance and the heterogeneous nature of ultrasound images. Participants implemented various strategies to tackle these challenges, many of which built upon the approaches discussed in the segmentation task.

\textbf{Leveraging Segmentation and Detection Results.}  
In several cases, classification was directly integrated with segmentation and detection tasks to improve overall performance. Segmentation outputs were frequently used to identify regions of interest (ROIs) that were then classified. This approach allowed models to focus on tumor-specific areas, reducing false positives and improving classification accuracy.

\textbf{Model Architectures and Customization.}  
While many teams employed advanced architectures such as ResNet and DenseNet, their strategies for classification mirrored the customization efforts discussed in the segmentation task. These included using smaller patch sizes to focus on localized features, applying residual connections to enhance feature extraction, and leveraging ensemble models for robust predictions.

\textbf{Addressing Class Imbalance.}  
As with segmentation, addressing class imbalance was critical for classification. Strategies such as focal loss and balanced sampling were commonly used to ensure that the model effectively learned patterns for both benign and malignant lesions. Similar to segmentation, some teams prioritized sampling of foreground regions or underrepresented classes during training.

\textbf{Data Augmentation and Preprocessing.}  
Data augmentation techniques, including flipping, rotation, and intensity normalization, were extensively applied, similar to the segmentation task. These augmentations increased data variability and reduced overfitting. Additionally, preprocessing focused on extracting patches from regions of highest interest, ensuring the classifier concentrated on areas most relevant for malignancy prediction.

\textbf{Hierarchical and Multi-step Classification.}  
A unique strategy in classification was the use of hierarchical or multi-step workflows. In these approaches, coarse classification was first applied to identify tumor-containing regions, followed by a second-stage classifier to differentiate between benign and malignant lesions. This hierarchical design minimized the impact of irrelevant background and enhanced focus on tumor-specific features.

\textbf{Post-Processing and Ensemble Learning.}  
Ensemble learning played a key role in improving classification performance. Similar to segmentation, multiple models trained with different architectures or initialization settings were combined to reduce prediction variance and enhance stability. Post-processing techniques, such as thresholding and confidence-based refinement, were employed to filter out low-confidence predictions and improve reliability.

\textbf{Feature-Based Classification.}  
Some teams supplemented deep learning methods with traditional feature-based approaches. Tumor-specific features, such as volume or shape, were extracted from segmented regions and used alongside learned features for malignancy prediction. This hybrid strategy added interpretability to the classification results and provided a secondary validation of deep learning predictions.

By employing strategies that built upon those used in segmentation while introducing unique hierarchical workflows and feature-based enhancements, participants effectively tackled the challenges of classification in 3D breast ultrasound data. These approaches highlight the importance of integrating segmentation outputs, addressing class imbalance, and refining model predictions to achieve accurate and reliable breast cancer diagnosis.

\subsection{Improving Strategies for Detection}

Accurate lesion detection in 3D breast ultrasound is critical for identifying and localizing tumors within large volumetric data. Participants designed diverse approaches to tackle the challenges posed by the variability in lesion size and shape, as well as the sparsity of lesions compared to the large background regions.

A common strategy was to leverage segmentation results to identify candidate regions for detection. By extracting connected components from segmentation masks, models could focus on regions of interest, reducing false positives and improving efficiency. Many approaches integrated segmentation and detection tasks seamlessly, using the former to guide the latter.

To handle the diversity of lesion sizes, anchor-free detection architectures were frequently employed. These models avoided the constraints of predefined anchor boxes and dynamically adapted to the size and shape of lesions. Multi-scale designs were also incorporated to ensure sensitivity to both small and large tumors within the same framework.

Post-processing techniques played a vital role in refining detection results. Methods such as non-maximum suppression (NMS) and clustering algorithms were used to consolidate overlapping predictions and eliminate redundant detections. Additionally, ensemble methods combining multiple models helped improve stability and robustness by reducing individual model biases.

Addressing class imbalance was another critical challenge in detection tasks. Many teams adopted similar techniques as in segmentation, such as focal loss and balanced sampling, to ensure that sparse lesion regions were effectively detected without being overshadowed by the dominant background.

By building on segmentation outputs, adopting flexible detection architectures, and applying robust post-processing, participants demonstrated significant progress in 3D lesion detection.

\subsection{Limitations and Future Work}

While the Tumor Detection, Segmentation, and Classification Challenge on Automated 3D Breast Ultrasound (ABUS) 2023 provided a valuable platform for benchmarking algorithms, several limitations were identified. 

Firstly, some label noise existed in the training dataset, particularly in complex tumor boundary regions. Although this noise was corrected in the test set to ensure fairness, accurate annotation remains a challenge, especially for smaller or less distinct tumors. Addressing this issue in future datasets through multi-expert consensus or AI-assisted annotation is crucial.

Secondly, the dataset was limited to breast tumor cases and lacked representation of other abnormalities or healthy tissue. Future versions of the challenge will expand the dataset to include multiple tumor types, healthy cases, and diverse patient demographics to improve model generalizability.

Additionally, this challenge did not evaluate runtime efficiency or resource consumption, allowing computationally intensive methods like test-time augmentation. While this promotes algorithm performance, future challenges may consider runtime and hardware constraints to encourage clinically deployable solutions.

Finally, domain adaptation remains a challenge, as the dataset was sourced from limited imaging protocols. Incorporating multi-center data and unseen imaging distributions will be prioritized in future iterations to enhance robustness and clinical applicability.

\section{Conclusion}

The TDSC-ABUS 2023 Challenge advanced the development of automated methods for 3D breast ultrasound analysis, addressing segmentation, classification, and detection tasks. Participants tackled challenges like class imbalance and lesion variability with innovative solutions, showcasing the potential of state-of-the-art architectures and tailored strategies.

Many methods integrated segmentation, classification, and detection into cohesive pipelines, emphasizing the benefits of combining task-specific innovations with shared techniques like data augmentation and ensemble learning. These approaches improved performance across tasks while addressing the unique complexities of 3D volumetric data.

This challenge demonstrated significant progress toward reliable and efficient breast cancer diagnosis, offering valuable insights into leveraging automated tools for clinical applications. The findings provide a strong foundation for future advancements in breast ultrasound analysis and beyond.

\section*{Acknowledgments}
We would like to extend our deepest gratitude to Professor Sang Hyun Park from DGIST for his invaluable guidance and leadership throughout the entirety of this project. We also thank Soopil Kim for his significant contributions and support. Additionally, we are grateful to Daekyung Kim and Kyong Joon Lee from Monitor Corporation for their assistance in model training and performance evaluation, which greatly enhanced the quality of our work.

\bibliographystyle{elsarticle-harv}\biboptions{authoryear}
\bibliography{refs.bib}

\end{document}